\documentclass[tightenlines,floatfix,preprintnumbers,amsmath,superscriptaddress,
amssymb,amsbsy,nofootinbib,nobibnotes,twocolumn,longbibliography,prx,doi]{revtex4-1}

\usepackage{amsmath, amssymb, amsfonts,amsbsy,amsthm} 
\usepackage{mathtools}
\usepackage{graphicx,color} 
\usepackage{acronym}
\usepackage{units,subfigure} 
\usepackage{hyperref} 
\usepackage{bm}
\usepackage{times}
\usepackage{mathrsfs}
\usepackage{latexsym}
\usepackage{dsfont}
\usepackage{hyphenat} 
\usepackage{titlesec}
\usepackage{accents} 
\usepackage{enumitem} 
\usepackage{multirow}
\usepackage{xspace} 
\usepackage[normalem]{ulem} 

\definecolor{darkgreen}{rgb}{0.0, 0.5, 0.0}
\definecolor{purple}{rgb}{0.56, 0.0, 1.0}

\newcommand{\gw}{_\textsc{gw}}

\newcommand{\hatbf}[1]{\hat{\mathbf{#1}}}
\newcommand{\blue}[1]{#1}
\newcommand{\newtext}[1]{#1}

\newcommand{\yr}{\,\text{yr}}

\newcommand{\Hz}{\,\text{Hz}}
\newcommand{\mTVS}{\textsc{TVS}}
\newcommand{\mTV}{\textsc{TV}}

\newcommand{\mT}{\textsc{T}}
\newcommand{\mV}{\textsc{V}}
\newcommand{\mS}{\textsc{S}}
\newcommand{\OddsSN}{\mathcal{O}^\textsc{sig}_\textsc{n}}
\newcommand{\OddsGR}{\mathcal{O}^\textsc{ngr}_\textsc{gr}}
\newcommand{\SNRopt}{\text{SNR}_\textsc{opt}}
\newcommand{\SNR}{\text{SNR}}
\newcommand{\multinest}{\texttt{MultiNest}\xspace}

\begin{document}
\title{Polarization-based Tests of Gravity with the Stochastic Gravitational-Wave Background}

\author{Thomas Callister}
\email{tcallist@caltech.edu}
\affiliation{LIGO Laboratory, California Institute of Technology, Pasadena, CA 91125, USA}

\author{Sylvia Biscoveanu}
\affiliation{The Pennsylvania State University, University Park, PA 16802, USA}

\author{Nelson Christensen}
\affiliation{Carleton College, Northfield, MN 55057, USA}
\affiliation{Artemis, Universit\'{e} C\^{o}te d'Azur, Observatoire C\^{o}te d'Azur, CNRS, CS 34229, F-06304 Nice Cedex 4, France}

\author{Maximiliano Isi}
\affiliation{LIGO Laboratory, California Institute of Technology, Pasadena, CA 91125, USA}

\author{Andrew Matas}
\affiliation{University of Minnesota, Minneapolis, MN 55455, USA}

\author{Olivier Minazzoli}
\affiliation{Artemis, Universit\'{e} C\^{o}te d'Azur, Observatoire C\^{o}te d'Azur, CNRS, CS 34229, F-06304 Nice Cedex 4, France}
\affiliation{Centre Scientifique de Monaco, 8 Quai Antoine 1er, MC 98000, Monaco}

\author{Tania Regimbau}
\affiliation{Artemis, Universit\'{e} C\^{o}te d'Azur, Observatoire C\^{o}te d'Azur, CNRS, CS 34229, F-06304 Nice Cedex 4, France}

\author{Mairi Sakellariadou}
\affiliation{King's College London, University of London, Strand, London WC2R 2LS, UK}

\author{Jay Tasson}
\affiliation{Carleton College, Northfield, MN 55057, USA}

\author{Eric Thrane}
\affiliation{School of Physics and Astronomy, Monash University, Clayton, Victoria 3800, Australia}
\affiliation{OzGrav: The ARC Centre of Excellence for Gravitational-Wave Discovery, Hawthorn, Victoria 3122, Australia}

\date{\today}

\begin{abstract}

The direct observation of gravitational waves with Advanced LIGO and Advanced Virgo offers novel opportunities to test general relativity in strong-field, highly dynamical regimes.
One such opportunity is the measurement of gravitational-wave polarizations.
While general relativity predicts only two tensor gravitational-wave polarizations, general metric theories of gravity allow for up to four additional vector and scalar modes.
The detection of these alternative polarizations would represent a clear violation of general relativity.
The LIGO-Virgo detection of the binary black hole merger GW170814 has recently offered the first direct constraints on the polarization of gravitational waves.
The current generation of ground-based detectors, however, is limited in its ability to sensitively determine the polarization content of transient gravitational-wave signals.
Observation of the stochastic gravitational-wave background, in contrast, offers a means of directly measuring generic gravitational-wave polarizations.
The stochastic background, arising from the superposition of many individually unresolvable gravitational-wave signals, may be detectable by Advanced LIGO at design-sensitivity.
In this paper, we present a Bayesian method with which to detect and characterize the polarization of the stochastic background.
We explore prospects for estimating parameters of the background, and quantify the limits that Advanced LIGO can place on vector and scalar polarizations in the absence of a detection.
Finally, we investigate how the introduction of new terrestrial detectors like Advanced Virgo aid in our ability to detect or constrain alternative polarizations in the stochastic background.
We find that, although the addition of Advanced Virgo does not notably improve detection prospects, it may dramatically improve our ability to estimate the parameters of backgrounds of mixed polarization.

\end{abstract}

\maketitle


\section{INTRODUCTION}

\newtext{The recent Advanced LIGO-Virgo observations of coalescing binary black holes have initiated the era of gravitational-wave astronomy~\cite{Aasi2015,Acernese2015,LIGO2016a,Abbott2016b,Abbott2016,GW170104,GW170814}.}
Beyond their role as astrophysical messengers, gravitational waves offer unique opportunities to test gravity in previously unexplored regimes \cite{1973PhRvL..30..884E,1973PhRvD...8.3308E,Gair2013,Will:2014kxa}.
The direct detection of gravitational waves has already enabled novel experimental checks on general relativity, placing the best model-independent dynamical bound to date on the graviton mass and limiting deviations of post-Newtonian coefficients from their predicted values \cite{Abbott2016,Abbott2016c,TheLIGOScientificCollaboration2016a,deRham:2016nuf,GW170104}.

\begin{figure*}
  \includegraphics[width=1\textwidth]{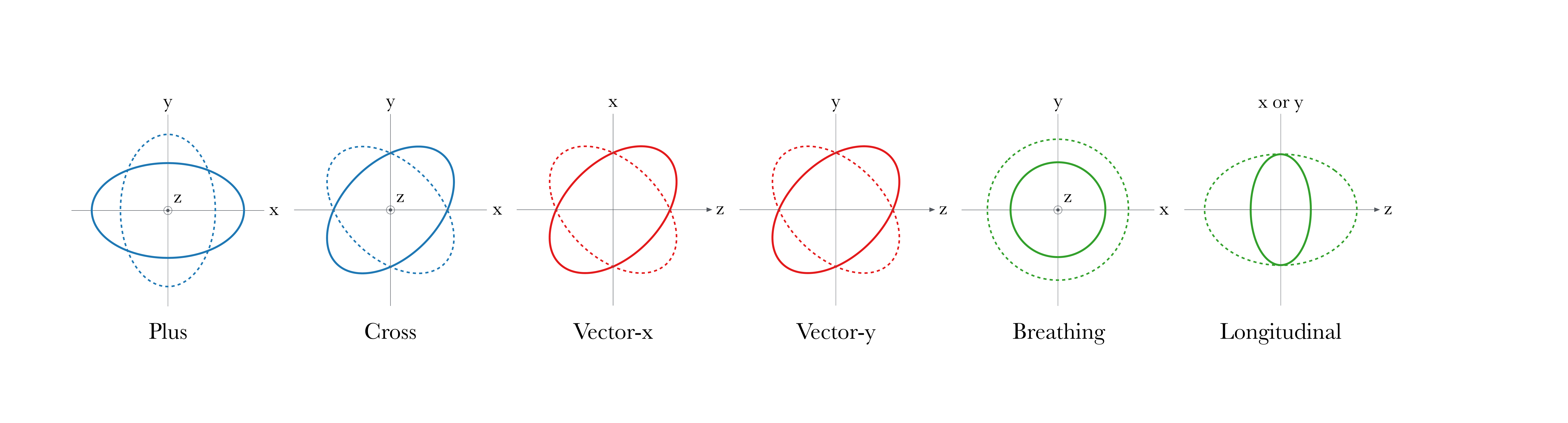}
  \onecolumngrid
  \caption{
  Deformation of a ring of freely-falling test particles under the six gravitational wave polarizations allowed in general metric theories of gravity.
  Each wave is assumed to propagate in the $z$-direction (out of the page for the plus, cross, and breathing modes; to the right for the vector-x, vector-y, and longitudinal modes).
  While general relativity allows only for two tensor polarizations (plus and cross), alternate theories allow for two vector (x and y) and/or two scalar (breathing and longitudinal) polarization modes.
  }
  \label{polarizations}
\end{figure*}

The measurement of gravitational-wave polarizations represents another avenue by which to test general relativity.
While general relativity allows for the existence of only two gravitational-wave polarizations (the tensor plus and cross modes),
general metric theories of gravity may allow for up to four additional polarizations: the $x$ and $y$ vector modes, and the breathing and longitudinal scalar modes \cite{1973PhRvL..30..884E,1973PhRvD...8.3308E,Will:2014kxa}.
The effects of all six polarizations on a ring of freely-falling test particles are shown in Fig. \ref{polarizations}.
The detection of these alternative polarization modes would represent a clear violation of general relativity, while their non-detection may serve to experimentally constrain extended theories of gravity.

\newtext{
Few experimental constraints exist on the polarization of gravitational waves \cite{Abbott2016c}.
Very recently, though, the simultaneous detection of GW170814 with the Advanced LIGO and Virgo detectors has allowed for the first direct study of a gravitational wave's polarization \cite{GW170814,GW170814_polarization}.
When analyzed with models assuming pure tensor, pure vector, and pure scalar polarization, GW170814 significantly favored the purely-tensor model over either alternative \cite{GW170814,GW170814_polarization}.
This result represents a significant first step in polarization-based tests of gravity.
Further tests with additional detectors, though, will be needed to sensitively test general relativity and its alternatives.

In particular, many alternate theories of gravity yield signals of \textit{mixed} polarizations, yielding vector and/or scalar modes in addition to standard tensor polarizations.
When allowing generically for all six polarization modes, the three-detector Advanced LIGO-Virgo network is generally unable to distinguish the polarization of transient gravitational-wave signals, like those from binary black holes \cite{Will:2014kxa,Romano2016,Gair2013,Abbott2016c,Abbott2016,GW170814_polarization}.
First, two LIGO detectors are nearly co-oriented, leaving Advanced LIGO largely sensitive to only a single polarization mode \cite{Will:2014kxa,Abbott2016c,Abbott2016,GW170814}.
Second, even if the LIGO detectors were more favorably-oriented, a network of at least six detectors is generically required to uniquely determine the polarization content of a gravitational-wave transient \cite{PhysRevD.86.022004,Will:2014kxa,Gair2013}.
Some progress can be made via the construction of ``null-streams'' \cite{PhysRevD.86.022004}, but this method is infeasible at present without an independent measure of a gravitational wave's source position (such as an electromagnetic counterpart).
Future detectors like KAGRA \cite{Aso2013} or LIGO-India \cite{Iyer2011} will therefore be necessary to break existing degeneracies and confidently distinguish vector or scalar polarizations in gravitational-wave transients.
It should be noted that the scalar longitudinal and breathing modes induce perfectly-degenerate responses in quadrupolar detectors like Advanced LIGO and Virgo.
Thus a network quadrupolar detectors can at most measure \textit{five} independent polarization degrees of freedom \cite{Will:2014kxa,PhysRevD.86.022004,GW170814_polarization}.
}

Beyond the direct detection of binary coalescences, another target for current and future detectors is the observation of the astrophysical stochastic gravitational-wave background, formed via the superposition of all gravitational-wave sources that are too weak or too distant to individually resolve \cite{Romano2016, Rosado2011,Zhu2011,Wu2012,Zhu2013,TheLIGOScientificCollaboration2016b}.
Although the strength of the background remains highly uncertain, it may be detected by Advanced LIGO in as few as \blue{two} years of coincident observation at design-sensitivity \cite{TheLIGOScientificCollaboration2016,TheLIGOScientificCollaboration2016b,2017PhRvL.118l1102A}.
\newtext{
Unlike direct searches for binary black holes, Advanced LIGO searches for long-lived sources like the stochastic background and rotating neutron stars \cite{2015CQGra..32x3001B,Nishizawa2009,Nishizawa2010,Nishizawa2013,Romano2016,2015PhRvD..91h2002I,Max} \textit{are} currently capable of directly measuring generic gravitational-wave polarizations without the introduction of additional detectors or identification of an electromagnetic counterpart.
}
The observation of the stochastic background would therefore enable novel checks on general relativity not possible with transient searches using the current generation of gravitational-wave detectors.

In this paper, we explore the means by which Advanced LIGO can detect and identify alternative polarizations in the stochastic background.
First, in Sect. \ref{theory}, we consider possible theorized sources which might produce a background of alternative polarizations.
We note, though, that stochastic searches are largely unmodeled, requiring few assumptions about potential sources or theories giving rise to alternative polarization modes (see, however, Sect. \ref{brokenSection}).

In Sect. \ref{nonGRBackgrounds} we discuss the tools used for detecting the stochastic background and compare the efficacy of standard methods with those optimized for alternative polarizations.
In Sect. \ref{bayesianSearch} we then propose a Bayesian method with which to both detect generically-polarized backgrounds and determine if alternative polarization modes are present.
Next, in Sect. \ref{peSection} we explore prospects for estimating the polarization content of the stochastic background.
We quantify the limits that Advanced LIGO can place on the presence of alternative polarizations in the stochastic background, limits which may be translated into constraints on specific alternative theories of gravity.

As new detectors are brought online in the coming years, searches for alternative polarizations in the stochastic background will become ever more sensitive.
In both Sects. \ref{bayesianSearch} and \ref{peSection}, we therefore investigate how the addition of Advanced Virgo improves our ability to detect or constrain backgrounds of alternative polarizations.
Finally, in Sect. \ref{brokenSection} we ask if our proposed search is robust against unexpectedly complex backgrounds of standard tensor polarizations.

\section{EXTENDED THEORIES OF GRAVITY AND ALTERNATIVE POLARIZATION MODES}
\label{theory}

Searches for the stochastic background are largely unmodeled, making minimal assumptions about the source of a measured background.
Nevertheless, it is interesting to consider which sources might give rise to a detectable background of alternative polarization modes.
In this section we briefly consider several possibilities that have been proposed in the literature.
We will focus mainly on scalar-tensor theories, which predict both tensor and scalar-polarized gravitational waves \cite{PhysRev.124.925}.
Our discussion below is not meant to be exhaustive; there may well exist additional sources that can give rise to backgrounds of extra polarization modes.
In particular, we do not discuss possible sources of vector modes, predicted by various alternative theories of gravity (see Ref. \cite{2015CQGra..32x3001B} and references therein).
Note that, while advanced detectors may not be sensitive to the sources described below, these sources may become increasingly relevant for third generation detectors (or beyond).

Core-collapse supernovae (CCSNe) represent one potential source of scalar gravitational waves.
Although spherically-symmetric stellar collapses do not radiate gravitational waves in general relativity, they do emit scalar breathing modes in canonical scalar-tensor theories.
While the direct observation of gravitational waves from CCSNe is expected to place strong constraints on scalar-tensor theories \cite{Gerosa2016}, only supernovae within the Milky Way are likely to be directly detectable using current instruments \cite{Gossan2016,Abbott2016e}.
Such events are rare, occurring at a rate between $(0.6-10.5)\times10^{-2}\yr^{-1}$ \cite{Adams2013}.
The stochastic gravitational-wave background, on the other hand, is dominated by distant undetected sources, and so in principle it is possible that a CCSNe background of breathing modes could be detected before the observation of a single Galactic supernova \cite{Crocker2015,2017PhRvD..95f3015C}.
However, realistic simulations of monopole emission from CCSNe predict only weak scalar emission \cite{Gerosa2016}.
Nevertheless, certain extreme phenomenological supernovae models predict gravitational radiation many orders of magnitude stronger than in more conventional models \cite{Gossan2016}.
According to such models, CCSNe may contribute non-negligibly to the stochastic background.

Compact binary coalescences may also contribute to a stochastic background of scalar gravitational waves.
In many scalar-tensor theories, bodies may carry a ``scalar charge" that sources the emission of scalar gravitational waves \cite{Damour1992,Damour:1993hw}.
Monopole scalar radiation is suppressed due to conservation of scalar charge, but in a general scalar-tensor theory there is generally no conservation law suppressing dipole radiation.
Scalar dipole radiation from compact binaries is enhanced by a factor of $(v/c)^{-2}$ relative to ordinary quadrupole tensor radiation (where $v$ is the orbital velocity of the binary and $c$ the speed of light), and thus represents a potentially promising source of scalar gravitational waves.
Electromagnetic observations of binary neutron stars place stringent constraints on anomalous energy loss beyond that predicted by general relativity; these constraints may be translated into a strong limit on the presence of additional scalar-dipole radiation \cite{Freire2012,Antoniadis2013}.
Such limits, though, are strongly model-dependent, assuming \textit{a priori} only small deviations from general relativity.
Additionally, pure vacuum solutions like binary black holes are not necessarily subject to these constraints.
If, for example, the scalar field interacts with curvature only through a linear coupling to the Gauss-Bonnet term, scalar radiation is produced by binary black holes but not by binary neutron stars \cite{Yagi:2015oca,Barausse2016}.
Alternatively, binary black holes can avoid the no-hair theorem and obtain a scalar charge if moving through a time-dependent or spatially-varying background scalar field \cite{Horbatsch2012,Berti2013}.

A variety of exotic sources may generically contribute to stochastic backgrounds of alternative polarizations as well.
Cosmic strings, for instance, generically radiate alternative polarizations in extended theories of gravity and may therefore contribute extra polarization modes to the stochastic gravitational-wave background \cite{Damour1997,Olmez2010}.
Another potential source of stochastic backgrounds of alternative polarizations are the so-called ``bubble walls'' generated by first order phase transitions in the early Universe \cite{Maggiore2000,Caprini2008,Caprini2016}.
In scalar-tensor theories, bubbles are expected to produce strong monopolar emission \cite{Damour1992}.
Gravitational waves from bubbles are heavily redshifted, though, and today may have frequencies too low for Advanced LIGO to detect \cite{Caprini2008}.
Bubble walls may therefore be a more promising target for future space-based detectors like LISA than for current ground-based instruments.

Finally, we note that it is also possible for alternative polarizations to be generated more effectively from sources at very large distances. 
There are several ways in which this might occur.
First, modifications to the gravitational-wave dispersion relation can lead to mixing between different polarizations in vacuum (an effect analogous to neutrino oscillations). This can cause mixing between the usual tensor modes \cite{2016arXiv160801284T}, and also between tensor modes and other polarizations \cite{DeFelice:2013nba,2015PhRvD..91f2007N}.
Thus alternative polarizations can be generated during propagation, even if only tensor modes are produced at the source.
This effect would build with the distance to a given gravitational-wave source.
Such behavior is among the effects arising from generic Lorentz-violating theories of gravity \cite{Kostelecky:2003fs,Kostelecky:2016kfm}.
While birefringence and dispersion of the standard plus and cross modes have been explored observationally in this context \cite{Kostelecky:2016kfm,Yunes:2016jcc}, the phenomenological implications of additional polarization modes remain an open issue at present.
Secondly, in many alternative theories fundamental constants (such as Newton's constant $G$) are elevated to dynamical fields;
these fields may have behaved differently at earlier stages in the Universe's evolution \cite{damour:2012zr,damour:1993uq}.
As a consequence, local constraints on scalar emission may not apply to emission from remote sources.
Additionally, it is in principle possible for local sources to be affected by screening mechanisms that do not affect some remote sources \cite{2015PhR...568....1J}.
\section{STOCHASTIC BACKGROUNDS OF ALTERNATIVE POLARIZATIONS}
\label{nonGRBackgrounds}

\begin{figure*}
\centering
  \includegraphics[width=0.48\textwidth]{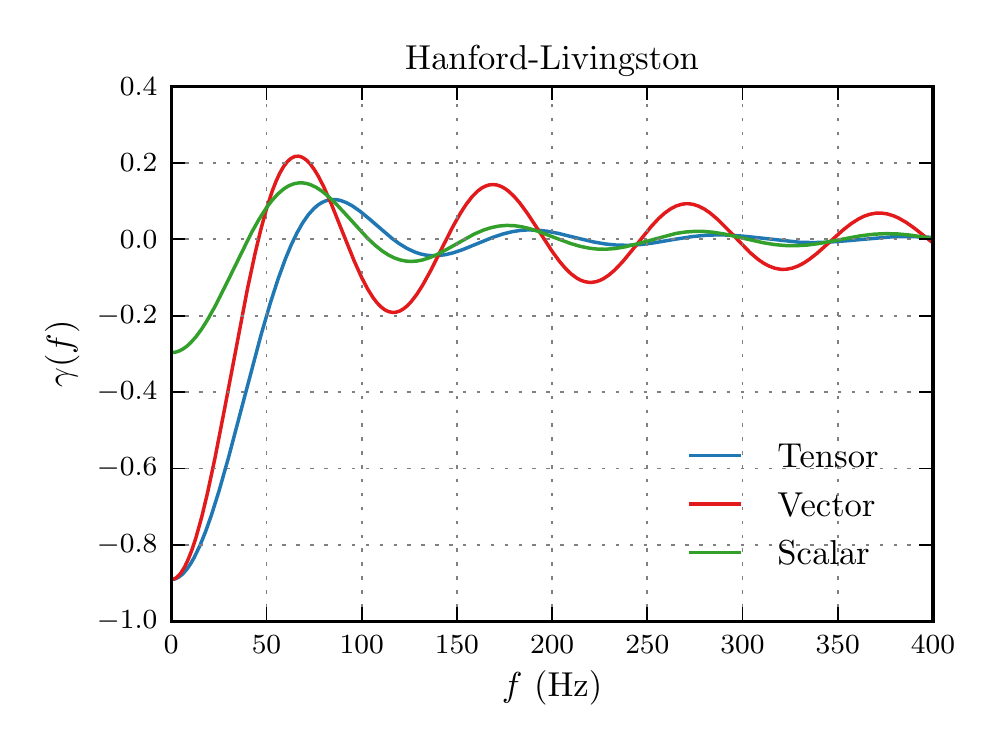}
  \includegraphics[width=0.48\textwidth]{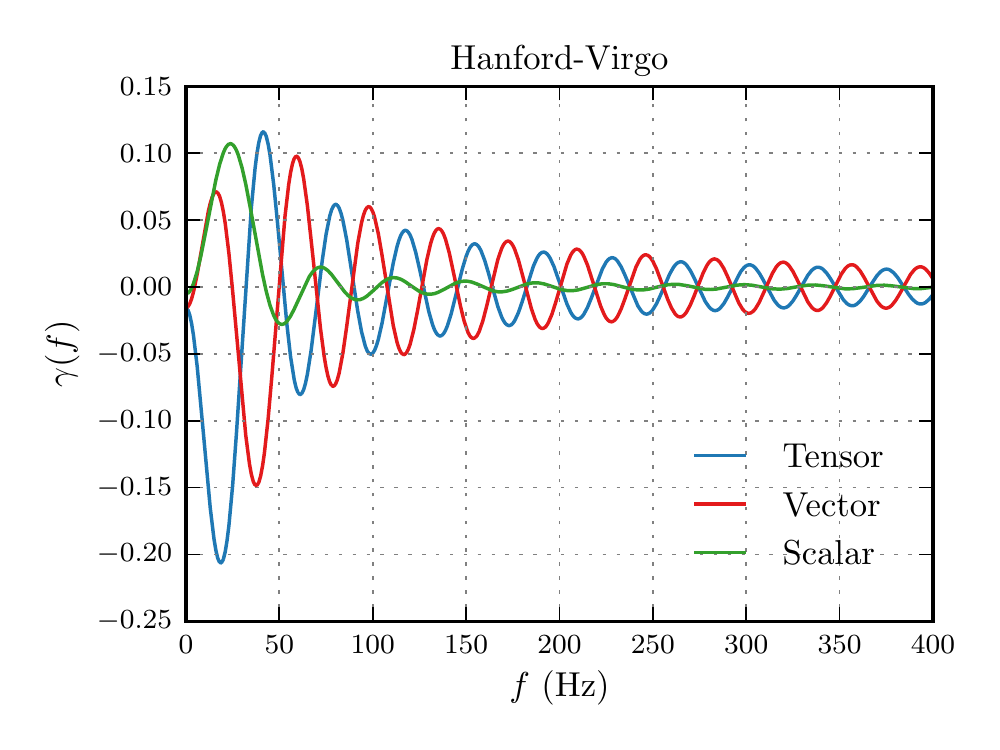}
  \caption{
Overlap reduction functions quantifying the sensitivities of the Hanford-Livingston (left) and Hanford-Virgo (right) baselines to isotropic backgrounds of tensor, vector, and scalar-polarized gravitational waves.
The distance between Hanford and Virgo is much larger than that between Hanford and Livingston; the Hanford-Virgo overlap reduction functions are therefore smaller in amplitude and more rapidly oscillatory.
  }
  \label{orfPlot}
\end{figure*}

The stochastic background introduces a weak, correlated signal into networks of gravitational-wave detectors.
Searches for the stochastic background therefore measure the cross-correlation
	\begin{equation}
	\hat C(f) \propto \tilde s_1^*(f) \tilde s_2(f)
	\end{equation}
between the strain $\tilde s_1(f)$ and $\tilde s_2(f)$ measured by pairs of detectors (see Ref. \cite{Romano2016} for a comprehensive review of stochastic background detection methods).

We will make several assumptions about the background.
First, we will assume that the stochastic background is isotropic, stationary, and Gaussian.
Second, we assume that there are no correlations between different tensor, vector, and scalar polarization modes.
We can therefore express the total measured cross-power $\langle \hat C(f)\rangle$ as a sum of three terms due to each polarization sector.
Finally, we assume that the tensor and vector sectors are individually unpolarized, with equal power in the tensor plus and cross modes and equal power in the vector-$x$ and vector-$y$ modes.
This follows from the fact that we expect gravitational-wave sources to be isotropically distributed and randomly oriented with respect to the Earth.
In contrast, we cannot assume that the scalar sector is unpolarized.
Scalar breathing and longitudinal modes cannot be rotated into one another via a coordinate transformation (as can the tensor plus and cross modes, for instance), and so source isotropy does not imply equal power in each scalar polarization.
However, the responses of the LIGO detectors to breathing and longitudinal modes are completely degenerate, and so Advanced LIGO is sensitive only to the total power in scalar modes rather than the individual energies in the breathing and longitudinal polarizations \cite{Will:2014kxa,Nishizawa2009}.

The above assumptions are not all equally justifiable, and may be broken by various alternative theories of gravity.
For instance, one should not expect an unpolarized background in any theory that includes parity-odd gravitational couplings, like Chern-Simons gravity \cite{Jackiw2003, Alexander2005, Alexander2008, Alexander2009}, even in the absence of non-tensorial modes \cite{Contaldi2008}.
Furthermore, different polarizations may not be statistically independent, as is the case for the breathing and longitudinal modes in linearized massive gravity \cite{IsiStein2017}.
Finally, we should expect a departure from isotropy in any theory violating Lorentz invariance, like those within the standard model extension framework \cite{Kostelecky:2003fs,Kostelecky:2016kfm,2016arXiv160801284T}.
These exceptions notwithstanding, for simplicity we will proceed under the assumptions listed above, leaving more generic cases for future work.

Under our assumptions, the measured cross-power due to the background is given by \cite{Romano2016,Allen1999,Nishizawa2009}
	\begin{equation}
	\label{crossPower}
	\langle \tilde s_1^*(f) \tilde s_2(f')\rangle = \delta(f-f') \gamma_a(f) H^a(f),	
	\end{equation}
where repeated indices denote summation over tensor, vector, and scalar modes ($a\in\{T,V,S\}$).
The overlap reduction functions $\gamma_a(f)$ quantify the sensitivity of detector pairs to isotropic backgrounds of each polarization \cite{Christensen1992,Nishizawa2009} (see Appendix \ref{orfAppendix} for details).
The functions $H^a(f)$, meanwhile, encode the spectral shape of the stochastic background within each polarization sector.

In the left side of Fig. \ref{orfPlot}, we show the overlap reduction functions for the Hanford-Livingston (H1-L1) Advanced LIGO network.
The overlap reduction functions are normalized such that $\gamma_T(f) = 1$ for coincident and coaligned detectors.
For the Advanced LIGO network, the tensor overlap reduction function has magnitude $|\gamma_T(0)| = 0.89$ at $f=0$, representing reduced sensitivity due to the separation and relative rotation of the H1 and L1 detectors.
Additionally, the H1-L1 tensor overlap reduction function decays rapidly to zero above $f\approx64\Hz$.
Standard Advanced LIGO searches for the stochastic background therefore have negligible sensitivity at frequencies above $\sim64\Hz$.

Relative to $\gamma_T(f)$, the H1-L1 vector overlap reduction function $\gamma_V(f)$ is of comparable magnitude at low frequencies, but remains non-negligible at frequencies above 64 Hz.
As a result, we will see that Advanced LIGO is in many cases more sensitive to vector-polarized backgrounds than standard tensor backgrounds.
The scalar overlap reduction function, meanwhile, is smallest in magnitude, with $|\gamma_S(0)|$ a factor of three small than $|\gamma_T(0)|$ and $|\gamma_V(0)|$.
Advanced LIGO is therefore least sensitive to scalar-polarized backgrounds.
This reflects a generic feature of quadrupole gravitational-wave detectors, which geometrically have a smaller response to scalar modes than to vector and tensor polarizations \cite{Max}.
For an extreme example of the opposite case, see pulsar timing arrays, which are orders of magnitude more sensitive to longitudinal polarizations than standard tensor-polarized signals \cite{2011PhRvD..83l3529A,2012PhRvD..85h2001C}.

For comparison, the right side of Fig. \ref{orfPlot} shows the overlap reduction functions for the Hanford-Virgo (H1-V1) baseline.
As the separation between Hanford and Virgo is much greater than that between Hanford and Livingston, the Hanford-Virgo overlap reduction functions are generally much smaller in amplitude and more rapidly oscillatory, translating into weaker sensitivity to the stochastic background.
Note, however, that the H1-V1 tensor overlap reduction function remains larger in amplitude than H1-L1's at frequencies $f\gtrsim200\Hz$, implying heightened relative sensitivity to tensor backgrounds at high frequencies \cite{2007CQGra..24S.639C}.

The functions $H^a(f)$ appearing in Eq. \eqref{crossPower} are theory-independent; they are observable quantities that can be directly measured in the detector frame.
Stochastic backgrounds are not conventionally described by $H(f)$, though, but by their gravitational-wave energy-density
 \cite{Allen1999}
	\begin{equation}
	\Omega(f) = \frac{1}{\rho_c} \frac{d\rho\gw(f)}{d\ln f},
	\end{equation}
defined as the fraction of the critical energy density $\rho_c = 3H_0^2 c^2/(8\pi G)$ contained in gravitational waves per logarithmic frequency interval $d\ln f$.
Here, $H_0$ is the Hubble constant and $G$ is Newton's constant.
Within general relativity, the background's energy-density is related to $H(f)$ via \cite{Allen1999}
	\begin{equation}
	\label{energyDensity}
	\Omega(f) = \frac{20\pi^2}{3 H_0^2} f^3 H(f).
	\end{equation}
Eq. \eqref{energyDensity} is a consequence of Isaacson's formula for the effective stress-energy of gravitational waves \cite{PhysRev.166.1272,Allen1999,IsiStein2017}.
Alternate theories of gravity, though, can predict different expressions for the stress-energy of gravitational-waves and hence different relationships between $H^a(f)$ and $\Omega^a(f)$ \cite{IsiStein2017}.
For ease of comparison to previous studies, we will use Eq. \eqref{energyDensity} to \textit{define} the canonical energy-density $\Omega^a(f)$ in polarization $a$.
If we allow Isaacson's formula to hold, then $\Omega^a(f)$ may be directly interpreted as a physical energy density.
If not, though, then $\Omega^a(f)$ can instead be understood as a function of the observable $H^a(f)$.

We will choose to normalize the cross-correlation statistic $\hat C(f)$ such that
	\begin{equation}
	\label{Y}
	\langle \hat C(f) \rangle = \gamma_a(f) \Omega^a(f).
	\end{equation}
Its variance is then \cite{Allen1999,Nishizawa2009}
	\begin{equation}
	\label{sigma2}
	\sigma^2(f) = \frac{1}{2 T df} \left(\frac{10\pi^2}{3H_0^2}\right)^2 f^6 P_1(f) P_2(f).
	\end{equation}
Here, $T$ is the total coincident observation time between detectors, $df$ is the frequency bin-width considered, and $P_i(f)$ is the noise power spectral density of detector $i$.
Note that the normalization of our cross-correlation measurement, with the overlap reduction functions appearing in $\langle\hat C(f)\rangle$ rather than $\sigma^2(f)$, differs from the convention normally adopted in the literature.
Standard stochastic searches typically define a statistic $\hat Y(f) \propto \tilde s^*_1(f) \tilde s_2(f)/\gamma_T(f)$, such that $\langle \hat Y(f) \rangle = \Omega^T(f)$ in the presence of a pure tensor background\cite{Callister2016,TheLIGOScientificCollaboration2016b,Aasi2014}.
Our choice of normalization, though, will prove more convenient when studying stochastic backgrounds of mixed gravitational-wave polarizations.
To emphasize this distinction, though, we denote our cross-power estimators by $\hat C(f)$, rather than the more common $\hat Y(f)$.

\begin{figure*}
  \centering
  \includegraphics[width=0.48\textwidth]{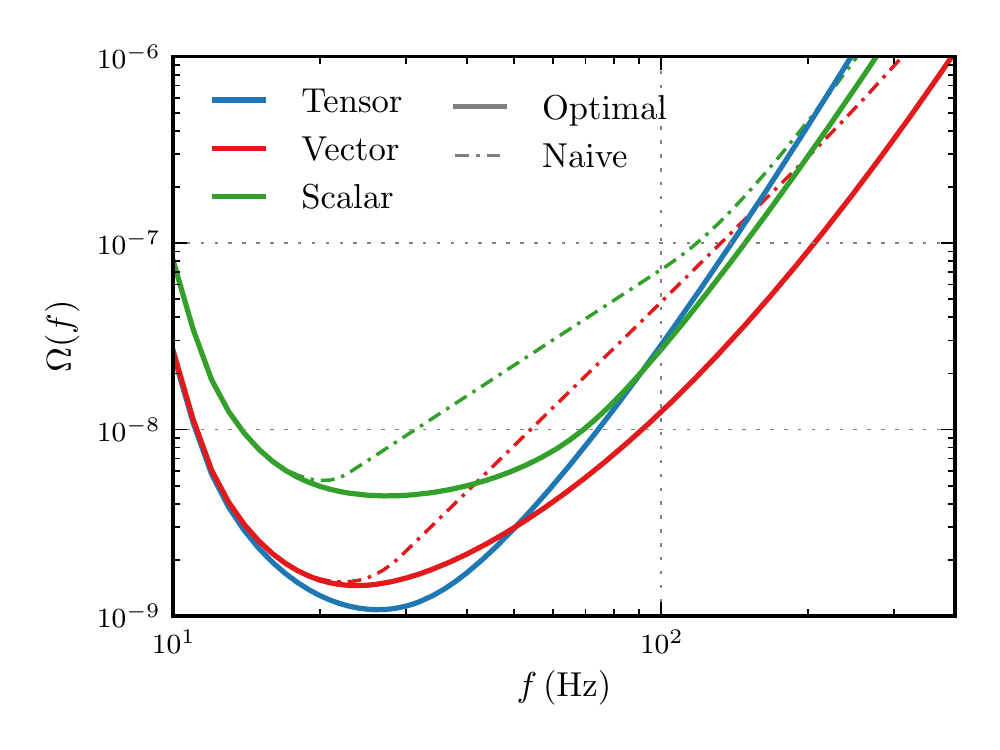}
  \includegraphics[width=0.48\textwidth]{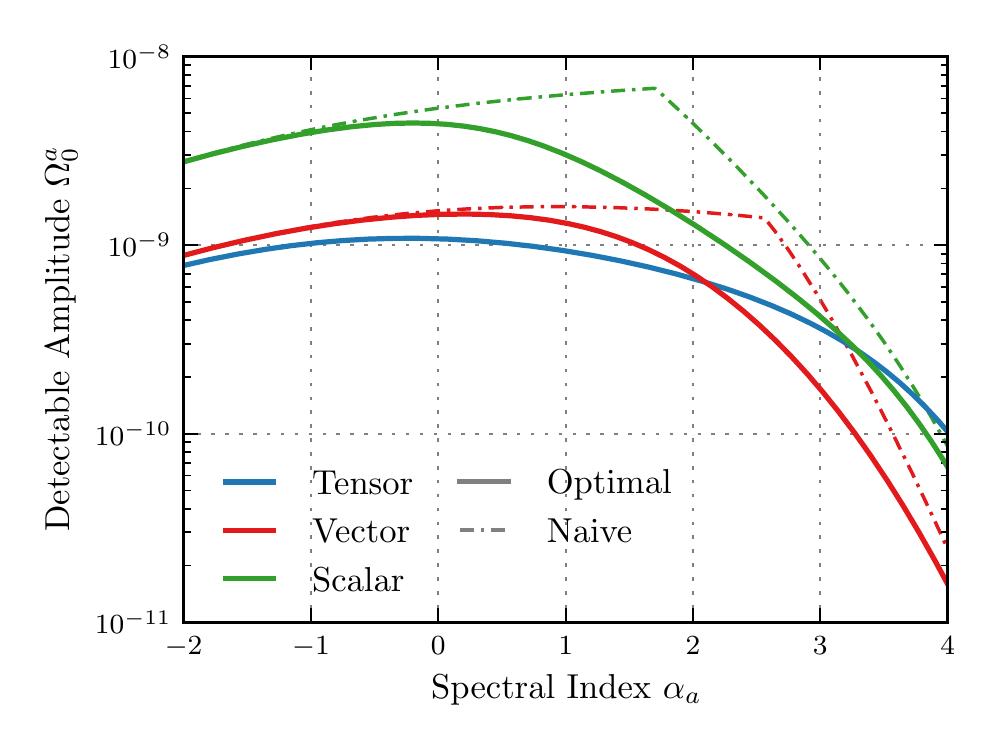}
  \caption{
\textit{Left}: PI curves showing the sensitivity of Advanced LIGO to stochastic backgrounds of tensor, vector, and scalar polarizations (solid blue, red, and green, respectively).
Power-law energy-density spectra [Eq. \eqref{powerLaw}] drawn tangent to the PI curves have expected $\langle\SNRopt\rangle=3$ after three years of observation at design-sensitivity.
Also shown are ``naive'' PI curves for vector and scalar backgrounds (dashed red and green) illustrating the sensitivity of existing search methods optimized only for tensor polarizations.
\textit{Right}: Minimum detectable background amplitudes ($\langle\SNRopt\rangle=3$ after three years of observation at design-sensitivity) as a function of spectral index $\alpha_a$.
For small and negative values of $\alpha_a$, Advanced LIGO is approximately equally sensitive to backgrounds of all three polarizations.
For large $\alpha_a$, Advanced LIGO is instead most sensitive to vector and scalar-polarized backgrounds.
The dashed curves show amplitudes detectable with existing ``naive'' methods.
The sensitivity loss between optimal and naive cases is negligible for $\alpha_a\lesssim0$, but becomes significant at moderate positive slopes (e.g. $\alpha_a\sim2$).
The kinks in the naive curves are due to biases incurred when recovering vector and scalar backgrounds with purely-tensor models; see the text for details.
  }
  \label{piCurves}
\end{figure*}

A spectrum of cross-correlation measurements $\hat C(f)$ may be combined to obtain a single broadband signal-to-noise ratio (SNR), given by
	\begin{equation}
	\label{snr}
	\text{SNR}^2 = \frac{ \bigl( \hat C \,|\, \gamma_a \Omega^a_M \bigr)^2 }
				{ \bigl( \gamma_b \Omega^b_M \,|\, \gamma_c \Omega^c_M \bigr) },
	\end{equation}
where we have defined the inner product	
	\begin{equation}
	\label{innerProduct}
	\left( A\,|\,B\right) = \left( \frac{3 H_0^2}{10\pi^2}\right)^2 2 T
		\int_0^\infty \frac{\tilde A^*(f) \tilde B(f)}{f^6 P_1(f) P_2(f)} df.
	\end{equation}
In Eq. \eqref{snr}, $\Omega^a_M(f)$ is our adopted model for the energy-density spectrum of the stochastic background.
The expected SNR is maximized when this model is equal to the background's true energy-density spectrum.
The resulting optimal SNR is given by
	\begin{equation}
	\label{optimalSNR}
	\text{SNR}^2_\textsc{opt} = ( \gamma_a\Omega^a \,|\, \gamma_b\Omega^b )
	\end{equation}
(see Appendix \ref{snrAppendix} for details).

Conventionally, stochastic energy-density spectra are modeled as power laws, such that
	\begin{equation}
	\label{powerLaw}
	\Omega^a_M(f) = \Omega^a_0 \left(\frac{f}{f_0}\right)^{\alpha_a},
	\end{equation}
where $\Omega^a_0$ is the background's amplitude at a reference frequency $f_0$ and $\alpha_a$ is its spectral index (or slope) \cite{Allen1999,TheLIGOScientificCollaboration2016b,Aasi2014}.
The predicted tensor stochastic background from compact binary coalescences, for instance, is well-modeled by a power law of slope $\alpha_T=2/3$ in the sensitivity band of Advanced LIGO \cite{Callister2016}.
For reference, slopes of $\alpha=0$ and $\alpha=3$ correspond to scale-invariant energy and strain spectra, respectively.
While we will largely assume power-law models in our analysis, in Sect. \ref{brokenSection} we will explore the potential consequences if this assumption is in fact incorrect (as would be the case, for instance, for a background of unexpectedly massive binary black holes \cite{Callister2016}).
Throughout this paper we will use the reference frequency $f_0=\blue{25\Hz}$.

With the above formalism in hand, we can quantify Advanced LIGO's sensitivity to stochastic backgrounds of alternative polarizations.
Plotted on the left side of Fig. \ref{piCurves} are power-law integrated (PI) curves representing Advanced LIGO's optimal sensitivity to power-law backgrounds of pure tensor (solid blue), vector (solid red), and scalar (solid green) modes \cite{Thrane2013a}.
The PI curves are defined such that a power-law spectrum drawn tangent to the PI curve will be marginally detectable with $\langle\SNRopt\rangle = \blue{3}$ after \blue{three} years of observation with design-sensitivity Advanced LIGO.
In general, energy-density spectra lying above and below the PI curves are expected to have optimal SNRs greater and less than $3$, respectively.
In the right side of Fig. \ref{piCurves}, meanwhile, the solid curves trace the power-law amplitudes required for marginal detection ($\langle\SNRopt\rangle = 3$ after three years of observation) as a function of spectral index.
Incidentally, the left and right-hand subplots of Fig. \ref{piCurves} are Legendre transforms of one another.

For spectral indices $\alpha_a\lesssim0$, Advanced LIGO is approximately equally sensitive to tensor and vector-polarized backgrounds, with reduced sensitivity to scalar signals.
When $\alpha_a=0$, for instance, the minimum optimally-detectable tensor and vector amplitudes are $\Omega^T_0 = \blue{1.1\times10^{-9}}$ and $\Omega^V_0=\blue{1.5\times10^{-9}}$, while the minimum detectable scalar amplitude is $\Omega^S_0 = \blue{4.4\times10^{-9}}$, a factor of three larger.
This relative sensitivity is due to the fact that the tensor and vector overlap reduction functions are of comparable magnitude at low frequencies, while the scalar overlap reduction function is reduced in size (see Fig. \ref{orfPlot}).

At high frequencies, on the other hand, Advanced LIGO's tensor overlap reduction function decays more rapidly than the vector and scalar overlap reduction functions.
As a result, Advanced LIGO is more sensitive to vector and scalar backgrounds of large, positive slope than to tensor backgrounds of similar spectral shape.
In Fig. \ref{piCurves}.a, for instance, the vector and scalar PI curves are seen to lie an order of magnitude below the tensor PI curve at frequencies above $f\sim300\Hz$.
The constraints that Advanced LIGO can place on positively-sloped vector and scalar backgrounds are therefore as much as an order of magnitude more stringent than those that can be placed on tensor backgrounds of similar slope.

We emphasize that the Advanced LIGO network's relative sensitivities to tensor, vector, and scalar-polarized backgrounds are due purely to its geometry, rather than properties of the backgrounds themselves.
If we were instead to consider the Hanford-Virgo baseline, for instance, the right-hand side of Fig. \ref{orfPlot} shows that at high frequencies the H1-V1 pair is least sensitive to scalar polarizations, whereas the H1-L1 baseline is least sensitive to tensor modes.

So far we have discussed only Advanced LIGO's \textit{optimal} sensitivity to stochastic backgrounds of alternative polarizations.
Existing stochastic searches, though, are \textit{not} optimized for such backgrounds, instead using models $\Omega^a_M(f)$ that allow only for tensor gravitational-wave polarizations.
The dashed curves in Fig. \ref{piCurves} illustrate Advanced LIGO's ``naive'' sensitivity to backgrounds of alternative polarizations when incorrectly assuming a purely-tensor model.
Note that the ``naive'' curves on the right side of Fig. \ref{piCurves} are not smooth, with sharp kinks at $\alpha_a\sim2$; more on this below.
The loss in sensitivity between the optimal and naive searches varies greatly with different spectral indices.
Sensitivity loss is relatively minimal for slopes $\alpha_a\lesssim0$.
When $\alpha_S=0$, for example, the minimum detectable scalar amplitude rises from $\Omega^S_0=\blue{4.4\times10^{-9}}$ in the optimal case to $\blue{5.3\times10^{-9}}$ in the naive case, an increase of \blue{20\%}.
Thus, a flat scalar background that is optimally detectable by Advanced LIGO may still be detected using existing techniques tailored to tensor polarizations.
The SNR penalty is more severe for stochastic backgrounds of moderate positive slope.
For $\alpha_S = 2$, Advanced LIGO can optimally detect a scalar background of amplitude $\Omega^S_0 = \blue{1.3\times10^{-9}}$, while existing methods would detect only a background of amplitude $\Omega^S_0 = \blue{4.4\times10^{-9}}$, a factor of \blue{3.4} larger.

\begin{figure}
  \centering
  \includegraphics[width=0.48\textwidth]{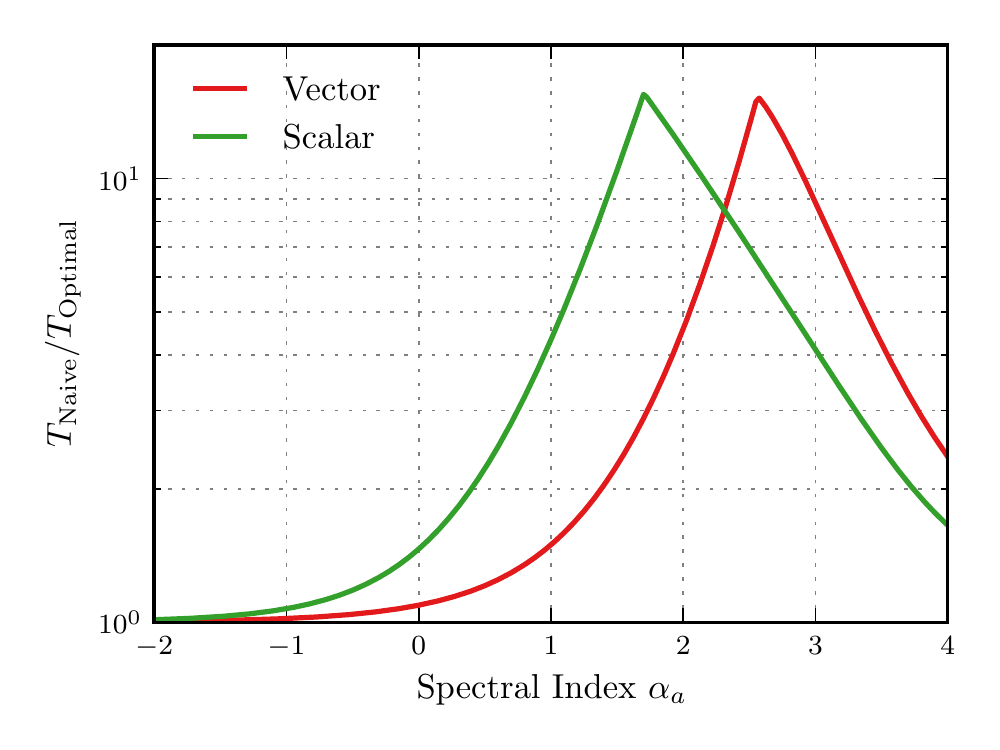}
  \caption{
The fractional increase in observing time required for Advanced LIGO to make a detection of vector (red) and scalar (green) backgrounds using existing search techniques, as a function of their spectral index $\alpha_a$.
The sharp kinks in each curve are due to biases incurred when fitting vector and scalar backgrounds with a purely-tensor model; see the text for details.
  }
  \label{timeIncrease}
\end{figure}

Since the SNR of the stochastic search accumulates only as $\text{SNR}\propto\sqrt{T}$, even a small decrease in sensitivity can result in a somewhat severe increase in the time required to make a detection.
To illustrate this, Fig. \ref{timeIncrease} shows the ratio $T_\text{Naive}/T_\text{Optimal}$ between the observing times required for Advanced LIGO to detect vector (red) and scalar (green) backgrounds using existing ``naive'' methods and optimal methods.
Although we noted above that existing methods incur little sensitivity loss to flat scalar backgrounds, the detection of such backgrounds would nevertheless require at least 50\% more observing time with existing searches.
Since the stochastic background is expected to be optimally detected only after several years, even a 50\% increase potentially translates into years of additional observation time, a requirement which may well stress standard experimental lifetimes and operational funding cycles.
Naive detection of a scalar background with $\alpha_S=2$, for comparison, would require nearly twelve times the observing time.

Figs. \ref{piCurves} and \ref{timeIncrease} both show conspicuous kinks occurring at $\alpha_S\approx1.75$ and $\alpha_V\approx2.5$.
These features are due to severe systematic parameter biases incurred when recovering vector and scalar backgrounds with a purely tensorial model.
For vector and scalar backgrounds of with $\alpha_a \gtrsim 3$, the best-fit slope $\alpha_T$ (which maximizes the recovered SNR) is biased towards large values.
Meanwhile, vector and scalar backgrounds with $\alpha_a \lesssim 1$ bias $\alpha_T$ in the opposite direction, towards smaller values.
The sharp kinks in Fig. \ref{piCurves} and \ref{timeIncrease} occur at the transition between these two regimes.
Such biases indicate another pitfall of existing search methods designed only for tensor-polarizations.
Even if a vector or scalar-polarized background is recovered with minimal SNR loss, without some independent confirmation we may remain entirely unaware that the detected background indeed violates general relativity (see Sect. \ref{bayesianSearch} below).
Furthermore, we would suffer from severe ``stealth bias,'' unknowingly recovering heavily-biased estimates of the amplitude and spectral index of the stochastic background \cite{2013PhRvD..87j2002V,2014PhRvD..89b2002V}.

\section{IDENTIFYING ALTERNATIVE POLARIZATIONS}
\label{bayesianSearch}

\begin{figure*}[ht]
  \centering
  \includegraphics[width=0.48\textwidth]{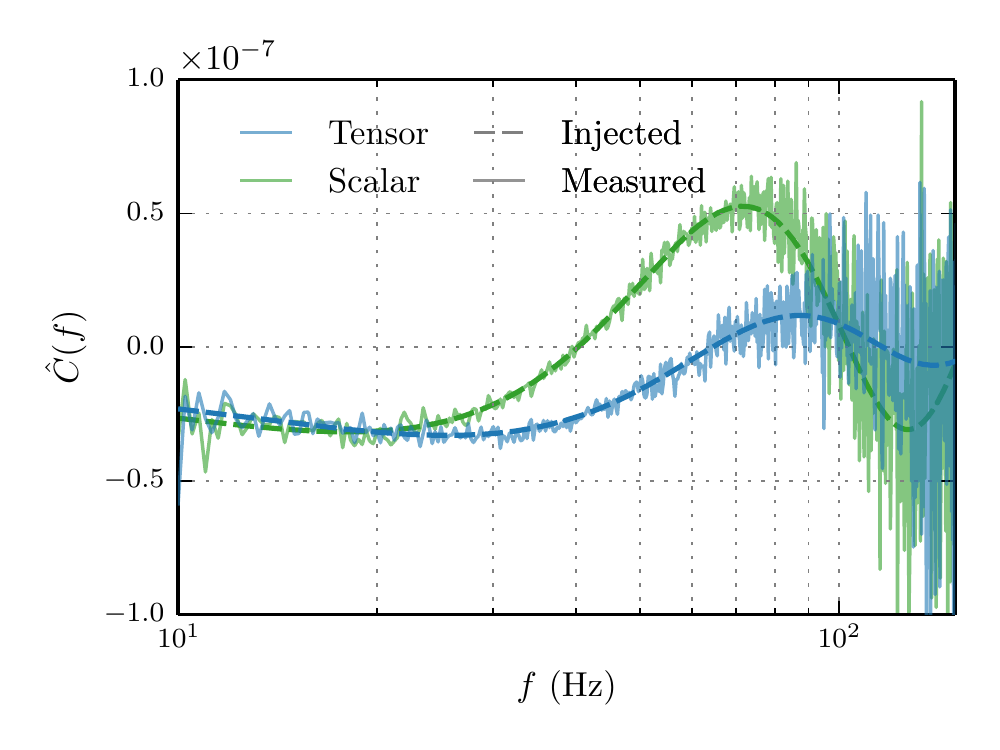}
  \includegraphics[width=0.48\textwidth]{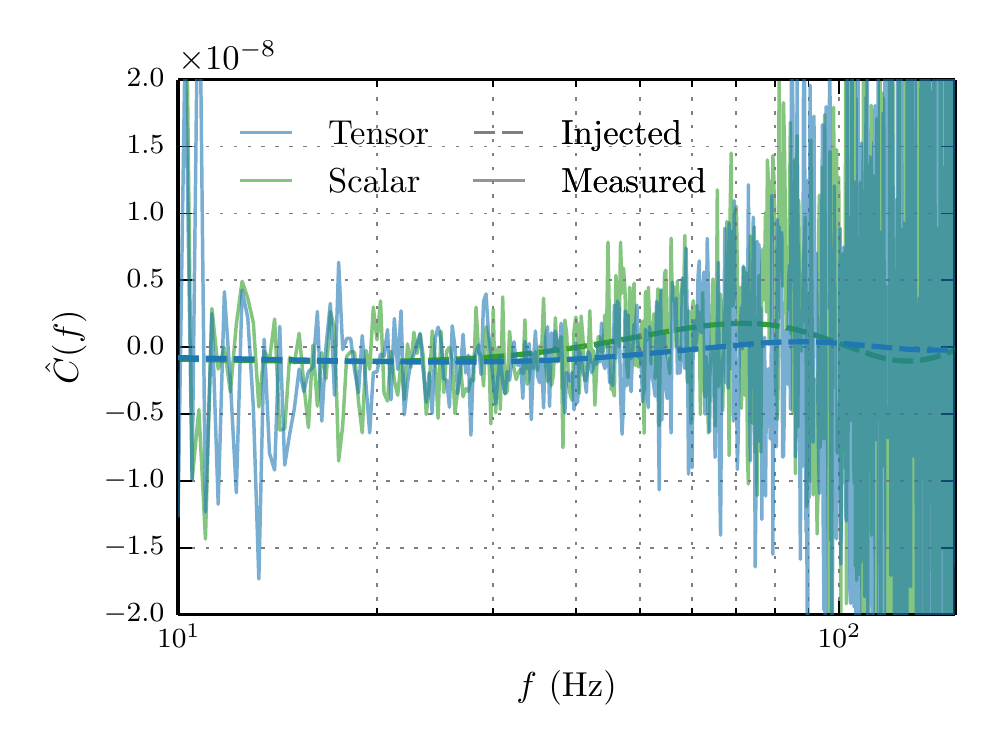}
  \caption{
\textit{Left}: Simulated cross-correlation measurements $\hat C(f)$ for purely tensor (blue) and purely scalar (green) stochastic backgrounds, recovered after \blue{three} years of observation with design-sensitivity Advanced LIGO.
The backgrounds shown have $\alpha_T = \alpha_S = 2/3$, and have amplitudes chosen such that each is detectable with $\langle \SNRopt\rangle = 150$.
The measured spectra each show distinct modulations characteristic of the tensor and scalar overlap reduction functions, allowing a clear identification of the polarization in each case.
\textit{Right}: Simulated recovery of weaker tensor and scalar backgrounds, detectable with $\langle \SNRopt \rangle = 5$ after three years of observation at design sensitivity.
While each background would be confidently detected by existing search techniques, the characteristic amplitude modulations and hence the polarization content of each simulated background are no longer evident.
  }
  \label{exampleBackgrounds}
\end{figure*}

We have seen in Sect. \ref{nonGRBackgrounds} that, even when using existing methods assuming only standard tensor polarizations, Advanced LIGO may still be capable of detecting a stochastic background of vector or scalar modes (albeit after potentially much longer observation times).
Detection is only the first of two hurdles, though.
Once the stochastic background has been detected, we will still need to establish whether it is entirely tensor-polarized, or if it contains vector or scalar-polarized gravitational waves.

Since tensor, vector, and scalar gravitational-wave polarizations each enter into cross-correlation measurements [Eq. \eqref{crossPower}] with unique overlap reduction functions, the polarization content of a detected stochastic background is in principle discernible from the spectral shape of $\hat C(f)$.
As an example, Fig. \ref{exampleBackgrounds} shows simulated cross-correlation measurements $\hat C(f)$ for both purely tensor (blue) and purely scalar-polarized (green) backgrounds after three years of observation with design-sensitivity Advanced LIGO.
The left-hand side shows simulated measurements of extremely strong backgrounds, with spectra $\Omega^T(f) = \blue{5\times10^{-8}} (f/f_0)^{2/3}$ and $\Omega^S(f) = \blue{1.8\times10^{-7}} (f/f_0)^{2/3}$; amplitudes are chosen such that each background has expected $\langle \SNRopt \rangle = \blue{150}$ after three years of observation.
The dashed curves trace the expectation values $\langle \hat C(f) \rangle$ of the cross-correlation spectra for each case, while the solid curves show a particular instantiation of measured values.
The alternating signs (positive or negative) of each spectrum are determined by the tensor and scalar overlap reduction functions, which have zero-crossings at different characteristic frequencies (see Fig. \ref{orfPlot}).
As a result, tensor and scalar-polarized signals each impart a unique shape to the cross-correlation spectra, offering a means of discriminating between the two cases.

As mentioned above, though, the backgrounds shown on the left side of Fig \ref{exampleBackgrounds} are unphysically loud, with $\SNRopt=\blue{152}$ and $\blue{148}$ for the simulated tensor and scalar backgrounds, respectively.
A tensor background of this amplitude would have been detectable with the standard isotropic search over Advanced LIGO's O1 observing run \cite{TheLIGOScientificCollaboration2016b}. 
Since stochastic searches accumulate SNR over time, the first detection of the stochastic background will necessarily be marginal; in this case the presence of alternative gravitational-wave polarizations would not be clear.
To demonstrate this, the right side of Fig. \ref{exampleBackgrounds} shows the simulated recovery of weaker tensor and scalar backgrounds of spectral shape $\Omega^T(f) = \blue{1.7\times10^{-9}} (f/f_0)^{2/3}$ and $\Omega^S(f) = \blue{6.1\times10^{-9}}(f/f_0)^{2/3}$, again after three years of observation with Advanced LIGO.
These amplitudes correspond to expected $\langle \SNRopt\rangle = 5$ after three years.
While Advanced LIGO would still make a very confident detection of each background, with $\SNRopt=\blue{6.7}$ and $\blue{7.8}$ for the simulated tensor and scalar cases, the backgrounds' polarization content is no longer obvious.

Interestingly, even when naively searching for purely-tensor polarized backgrounds, design-sensitivity Advanced LIGO still detect the ``quiet'' scalar example (on the right side of Fig. \ref{exampleBackgrounds}) with $\SNR=\blue{5.0}$.
When assuming \textit{a priori} that the stochastic background is purely tensor-polarized, any vector or scalar components detected with existing techniques may therefore be mistaken for ordinary tensor modes.
Not only would vector or scalar components fail to be identified, but, as discussed in Sect. \ref{nonGRBackgrounds}, they would heavily bias parameter estimation of the tensor energy-density spectrum.
If we wish to test general relativity with the stochastic background, we will therefore need to develop new tools in order to formally quantify the presence (or absence) of vector or scalar polarizations.
Additionally, while we have so far investigated only backgrounds of pure tensor, vector, or scalar polarization, most plausible alternative theories of gravity will predict backgrounds of \textit{mixed} polarization, with vector or scalar components in addition to a tensor component.
Any realistic approach must therefore be able to handle a stochastic background of completely generic polarization content.

Our approach will be to detect and classify the stochastic background using Bayesian model selection, adapting the method used in Ref. \cite{Max} to study the polarization content of continuous gravitational-wave sources.
First, we will define an odds ratio $\OddsSN$ between signal (SIG) and noise (N) hypotheses to determine if a stochastic background (of any polarization) has been observed.
Once a background is detected, we then construct a second odds ratio $\OddsGR$ to determine if the background contains only tensor polarization (GR hypothesis) or if there is evidence of alternative polarizations (the NGR hypothesis).
We describe the definition and construction of $\OddsSN$ and $\OddsGR$ in Appendix \ref{modelConstruction}.
Unlike existing detection methods that assume a pure tensor background, our scheme allows for the detection of generically-polarized stochastic backgrounds.
It encapsulates the optimal detection of tensor, vector, and scalar polarizations as described in Sect. \ref{nonGRBackgrounds}, and moreover enables the detection of more complex backgrounds of mixed polarization.

To compute the odds ratios $\OddsSN$ and $\OddsGR$, we use the \texttt{PyMultiNest} package \cite{Buchner2014}, which implements a Python wrapper for the nested sampling software \texttt{MultiNest} \cite{Feroz2008,Feroz2009,Feroz2013}.
\texttt{MultiNest}, an implementation of the nested sampling algorithm \cite{Skilling2004,Skilling2006}, is designed to efficiently evaluate Bayesian evidences [see Eq. \eqref{evidence}] in high-dimensional parameter spaces, even in the case of large and possibly-curving parameter degeneracies.
At little additional computational cost, \texttt{MultiNest} also returns posterior probabilities for each model parameter, allowing for parameter estimation in addition to model selection.
Details associated with running \texttt{MultiNest} are given in Appendix \ref{multinestAppendix}.

Our approach fundamentally differs from the strategy proposed by Nishizawa \textit{et al.} in Refs. \cite{Nishizawa2009,Nishizawa2010,Nishizawa2013}.
Nishizawa \textit{et al.} endeavor to separate and measure the background's tensor, vector, and scalar content within each frequency bin.
To solve for these three unknowns, three pairs of gravitational-wave detectors are required to break the degeneracy between polarizations.
A nice feature of this method is that it allows for the separation of polarization modes without the need for a parametrized model of the background's energy-density spectrum.
However, it has several drawbacks.
First, the Nishizawa \textit{et al.} component separation scheme requires at least three detectors.
Even then, this method is not very sensitive; covariances between polarization modes mean that only very loud backgrounds can be separated and independently detected with reasonable confidence.
Finally, Nishizawa \textit{et al.} are largely concerned with the \textit{detection} of a background, not the characterization of its spectral shape.
Ref. \cite{Nishizawa2013} does discuss parameter estimation on the stochastic background using a Fisher matrix formalism, but there are very well-known problems with this approach \cite{Vallisneri2008}.

Our method is more aggressive.
Rather than attempting to resolve the relative polarization content within each frequency bin, we assume a power-law model for the energy-density in each polarization mode (see Appendix \ref{modelConstruction}).
This allows us to confidently detect far weaker signals than the Nishizawa \textit{et al.} approach.
While this approach is potentially susceptible to bias if our model poorly fits the true background, it is a reasonable model for astrophysically plausible scenarios.
Even if the true background differs significantly from this model, we find in Sect. \ref{brokenSection} that potential bias is negligible.
Another advantage of our method is that it can be used with only two detectors and hence can be applied \textit{today}, rather than waiting for the construction of future gravitational-wave detectors.
Finally, in Sect. \ref{peSection}, we show that our Bayesian approach allows for full parameter estimation on the stochastic background, which properly takes into account the full degeneracies between background parameters (something a Fisher matrix analysis cannot do).

\subsection{Backgrounds of Single Polarizations}

\begin{figure*}
  \centering
  \includegraphics[width=0.48\textwidth]{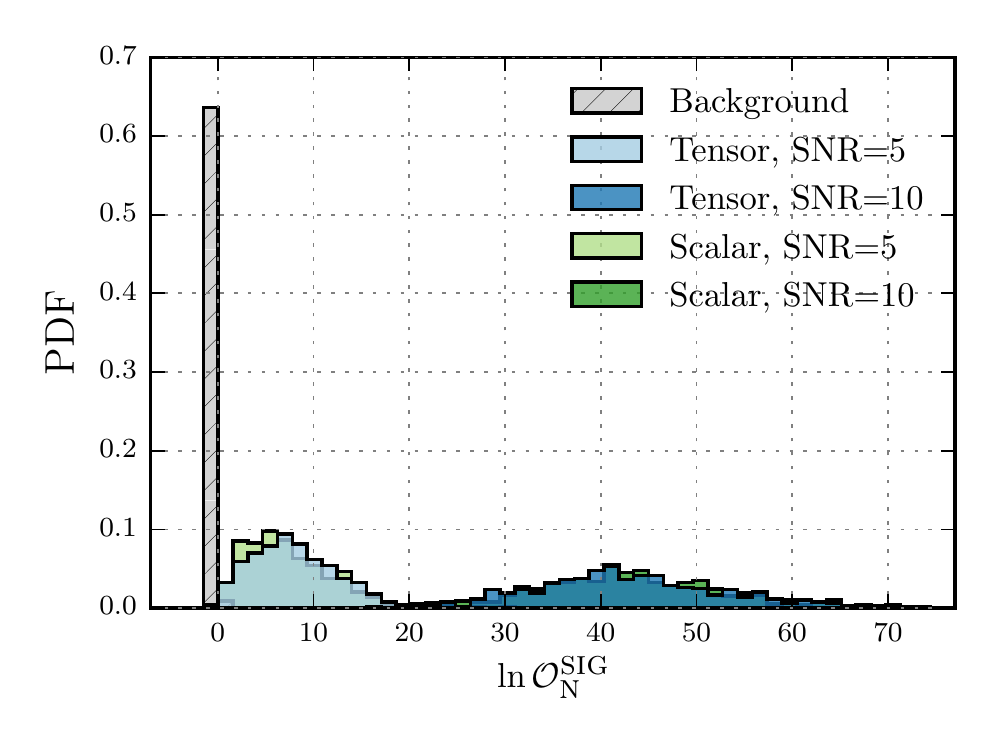}
  \includegraphics[width=0.48\textwidth]{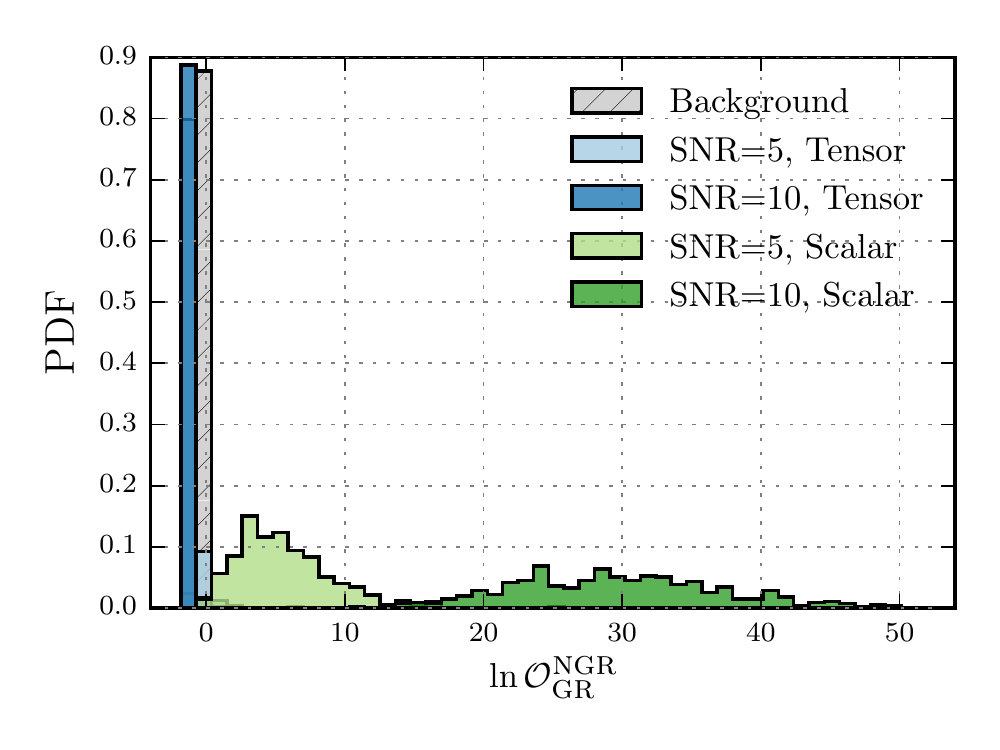}
  \caption{
\textit{Left}: Distributions of odd ratios $\OddsSN$ between signal and noise hypotheses for simulated observations of tensor (blue) and scalar (green) stochastic backgrounds of slope $\alpha=2/3$, assuming three years of observation with design-sensitivity Advanced LIGO.
We consider two different strengths for each polarization, corresponding to $\langle \SNRopt\rangle = 5$ and 10.
For each background strength, the tensor and scalar odds ratios lie nearly on top of one another.
Also shown is the background distribution of odds ratios obtained when observing pure Gaussian noise (hatched grey).
In the presence of a stochastic background, the recovered odds ratios grow as $\ln\OddsSN\propto\SNRopt^2$, showing increasingly large preference for the signal hypothesis.
\textit{Right}: Odds ratios $\OddsGR$ between NGR and GR hypotheses obtained for the same set of simulated Advanced LIGO observations.
In the presence of a tensor-polarized background, we recover narrow distributions of odds ratios centered at $\ln\OddsGR\approx\blue{-1.4}$, reflecting consistency with the GR hypothesis.
A scalar background, on the other hand, yields large positive odds ratios, correctly showing a strong preference for our NGR hypothesis.
}
\label{histograms}
\end{figure*}

As a first demonstration of this machinery, we explore the simple cases of purely tensor, vector, or scalar-polarized stochastic backgrounds.
Shown in Fig. \ref{histograms} are distributions of odds ratios $\OddsSN$ and $\OddsGR$ obtained for simulated observations of both tensor and scalar backgrounds, each of slope $\alpha=2/3$ (the characteristic slope of a tensor binary black hole background).
For each polarization, we consider two choices of amplitude, corresponding to $\langle \SNRopt\rangle = 5$ and $10$ after three years of observation with design-sensitivity Advanced LIGO.
For comparison, the hatched grey distributions show odds ratios obtained in the presence of pure Gaussian noise.

As seen in the left-hand side of Fig. \ref{histograms}, Gaussian noise yields a narrow odds ratio distribution centered at $\ln \OddsSN \approx \blue{-1.0}$ .
In contrast, the simulated observations of tensor and scalar backgrounds yield large, positive odds ratios, well-separated from Gaussian noise.
Note that the tensor and scalar distributions lie nearly on top of one another, as $\OddsSN$ depends primarily on the optimal SNR of a background and not its polarization content.

The right-hand side of Fig. \ref{histograms}, in turn, shows the odds ratios $\OddsGR$ quantifying the evidence for alternative polarization modes.
In the case of pure Gaussian noise, we again see a narrow distribution of odds ratios, centered at $\ln\OddsGR \approx \blue{-0.4}$.
In the absence of informative data, our analysis thus slightly favors the GR hypothesis.
This can be understood as a consequence of the implicit Bayesian ``Occam's factor," which penalizes the more complex NGR hypothesis over the simpler GR hypothesis.
Simulated observations of scalar backgrounds, in turn, yield large positive values for $\ln\OddsGR$, correctly preferencing the NGR hypothesis.
In contrast, pure tensor backgrounds yield negative $\ln\OddsGR$.
Interestingly, the recovered odds ratios do not grow increasingly negative with larger tensor amplitudes, but instead saturate at $\ln\OddsGR \approx \blue{-1.4}$.
This reflects the fact that a non-detection of vector or scalar polarizations can never strictly rule out their presence, but only place an upper limit on their amplitudes.
In other words, a strong detection of a pure tensor stochastic background cannot provide evidence \textit{for} the GR hypothesis, but at best only offers no evidence \textit{against} it.
This behavior is in part due to our choice of amplitude priors, which allow for finite but immeasurably small vector and scalar energy densities (see Appendix \ref{modelConstruction}).

\begin{figure*}
  \centering
  \includegraphics[width=0.48\textwidth]{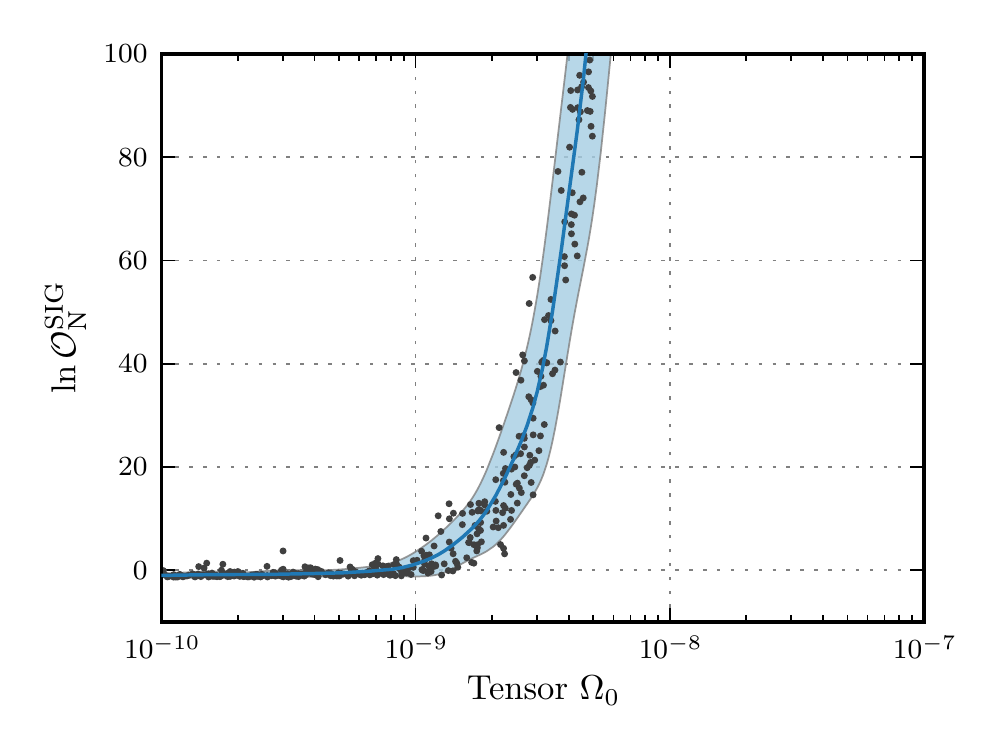}
  \includegraphics[width=0.48\textwidth]{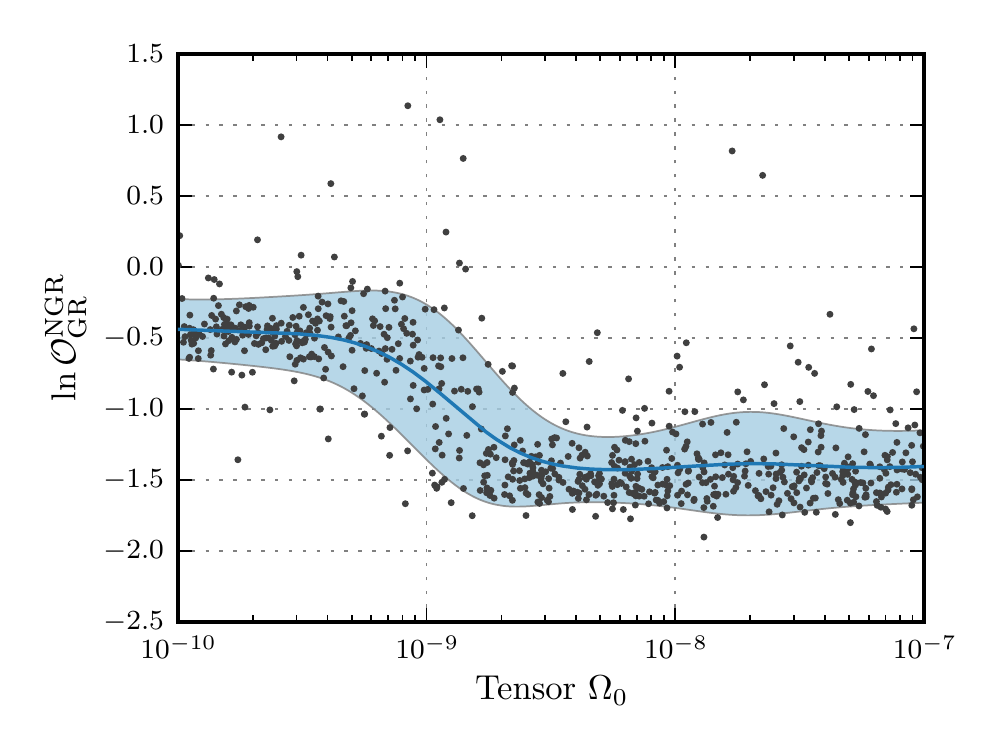} \\
  \includegraphics[width=0.48\textwidth]{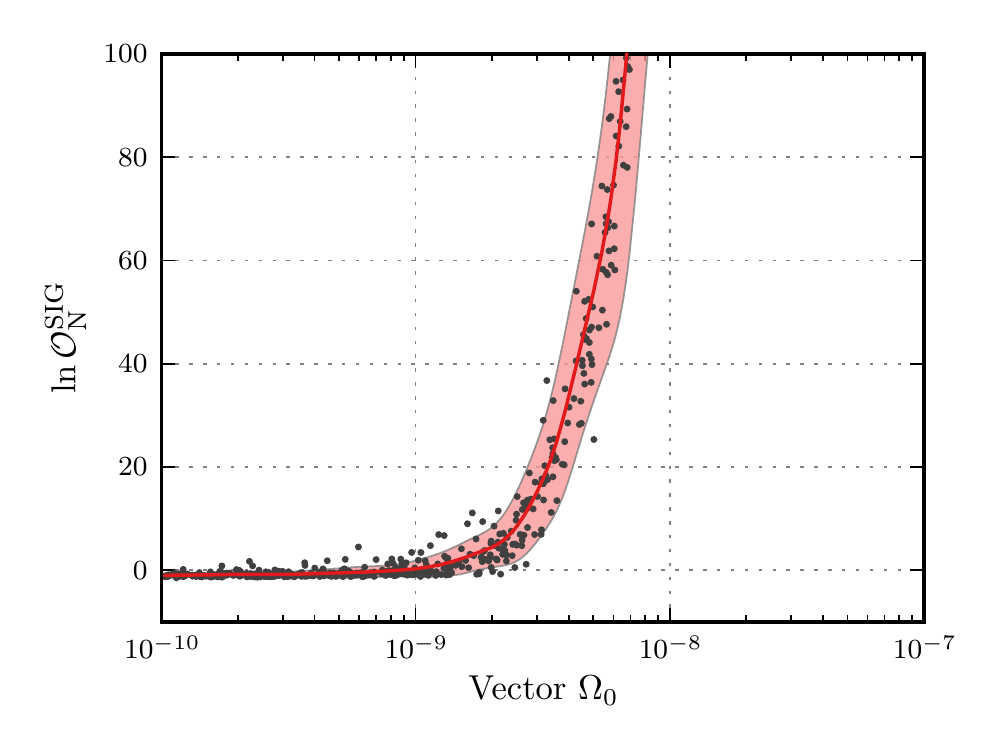}
  \includegraphics[width=0.48\textwidth]{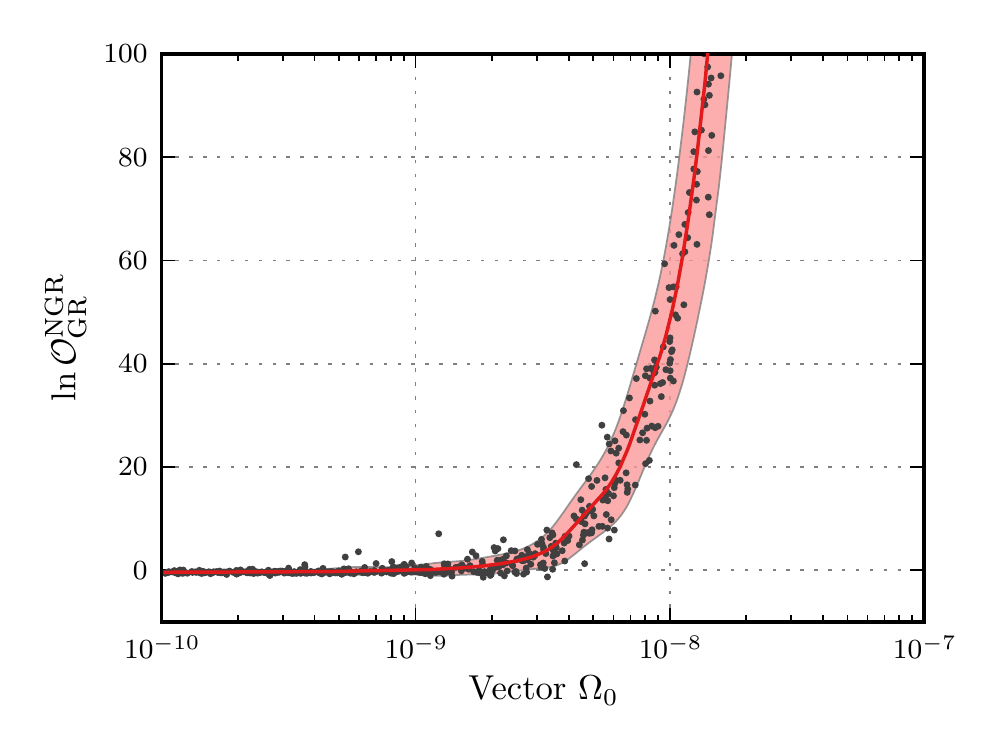} \\
  \includegraphics[width=0.48\textwidth]{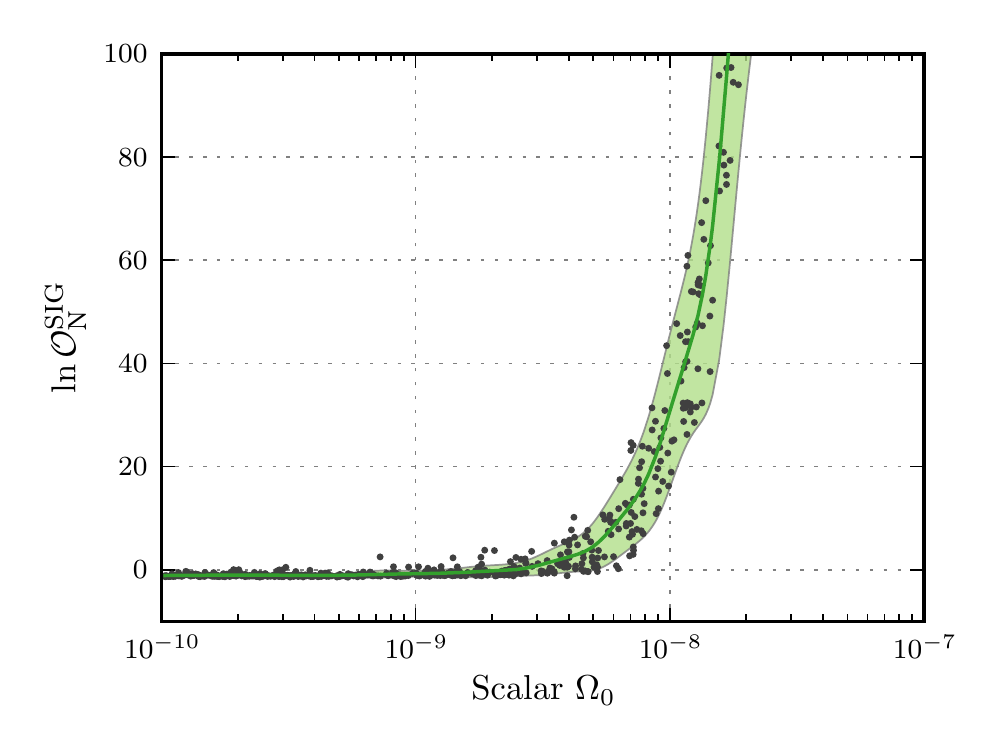}
  \includegraphics[width=0.48\textwidth]{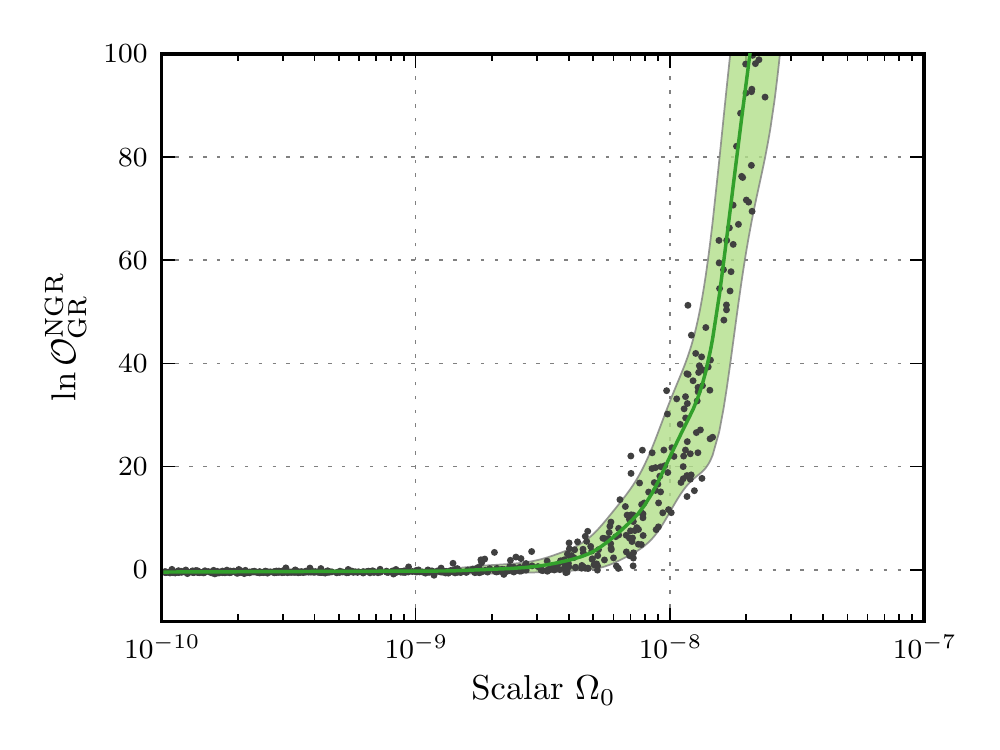}
  \caption{
Odds ratios $\OddsSN$ (left) and $\OddsGR$ (right) for simulated Advanced LIGO observations of purely tensor (blue), vector (red), and scalar (green) polarized stochastic backgrounds.
Within each plot, we show $\blue{750}$ simulated observations, with random log-amplitudes chosen uniformly over the range $-10<\log\Omega_0<-7$.
Black points mark the results from individual realizations, while the solid curves and shaded regions show the moving mean and standard deviations (smoothed with a Gaussian kernel) of these realizations.
For each polarization, $\log\OddsSN$ scales quadratically with the amplitude of the stochastic background.
Similarly, $\log\OddsGR$ scales quadratically with vector and scalar amplitude.
For tensor backgrounds, however, $\log\OddsGR$ instead saturates at approximately $\blue{-1.4}$.
  }
\label{oddsScatter}
\end{figure*}

Figure \ref{oddsScatter} illustrates more generally how $\OddsSN$ (left column) and $\OddsGR$ (right column) scale with the amplitudes of purely tensor (blue), vector (red), and scalar (green) stochastic backgrounds.
Black points mark odds ratios computed from individual realizations of simulated data, while the solid curves and shaded regions trace their smoothed mean and standard deviation.
We again see $\ln\OddsSN$ increasing monotonically with injected amplitude for all three polarizations.
Specifically, $\OddsSN$ depends inversely on the noise-hypothesis likelihood [defined by Eq. \eqref{noiseLikelihood}] and therefore scales as
	\begin{equation}
	\ln\OddsSN \propto \SNRopt^2.
	\end{equation}
As seen earlier in Fig. \ref{histograms}, $\ln\OddsGR$ saturates at $\blue{-1.4}$ for loud tensor backgrounds.
In the case of vector and scalar backgrounds, on the other hand, $\ln\OddsGR$ grows quadratically with increasing amplitude.
In particular, $\ln\OddsGR$ is proportional to the squared SNR of the \textit{residuals} between the observed $\hat C(f)$ and the best-fit tensor model.
We begin to see a strong preference for the NGR hypothesis when these residuals become statistically significant.

\subsection{Backgrounds of Mixed Polarization}

\begin{figure*}
  \centering
  \includegraphics[width=0.48\textwidth]{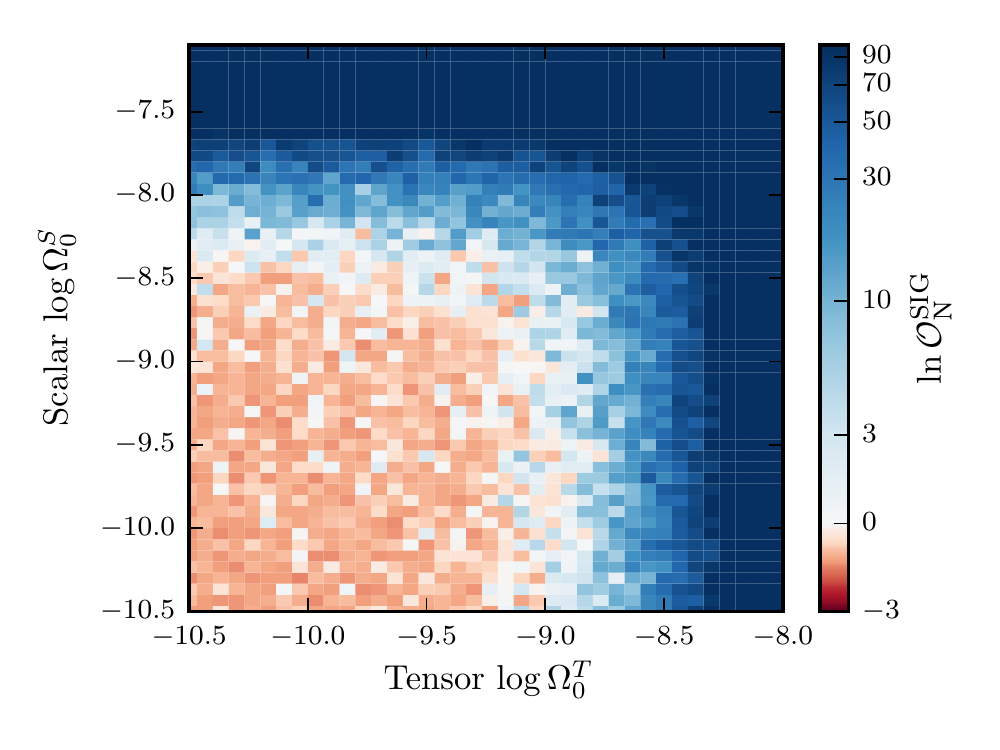}
  \includegraphics[width=0.48\textwidth]{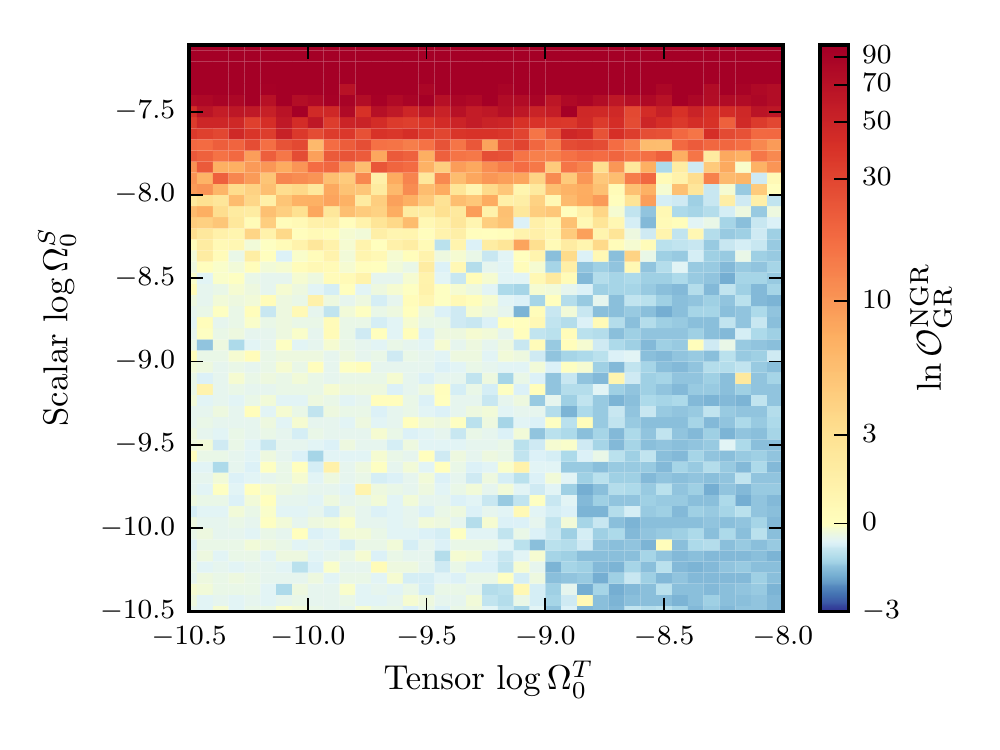}
  \caption{
Odds ratios for simulated Advanced LIGO measurements of stochastic backgrounds containing both tensor and scalar polarizations, assuming three years of observation at design sensitivity.
The tensor and scalar components have slopes $\alpha_T = 2/3$ and $\alpha_S = 0$, respectively.
\textit{Left}: Odds ratios between signal and noise hypotheses.
The observed values of $\OddsSN$ trace contours in total background energy.
Thus the detectability of a background depends largely on its total power, not its polarization content.
\textit{Right}: Odds ratios $\OddsGR$ between NGR and GR hypotheses.
Advanced LIGO would confidently identify the presence of the scalar background component when $\log\Omega^S_0 \gtrsim \blue{-7.9}$.
LIGO's sensitivity to the scalar component is nearly independent of the strength of the tensor component; the minimum identifiable scalar amplitude $\Omega^S_0$ rises only slightly with increasing $\Omega^T_0$.
  }
\label{STbayes}
\end{figure*}

\begin{figure*}
  \centering
  \includegraphics[width=0.48\textwidth]{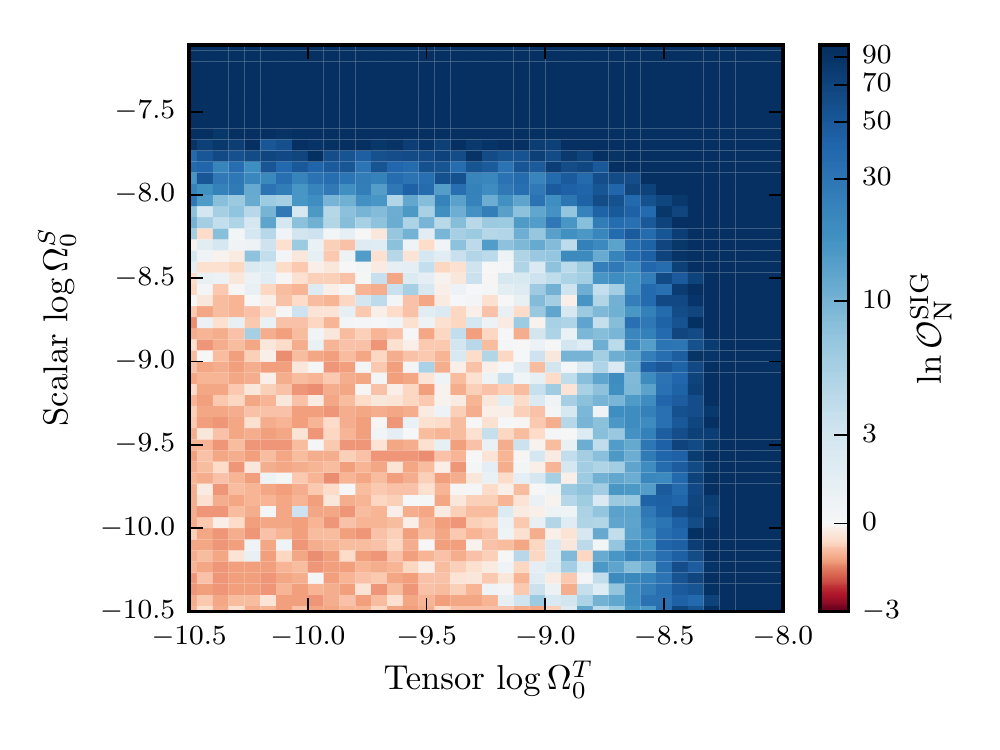}
  \includegraphics[width=0.48\textwidth]{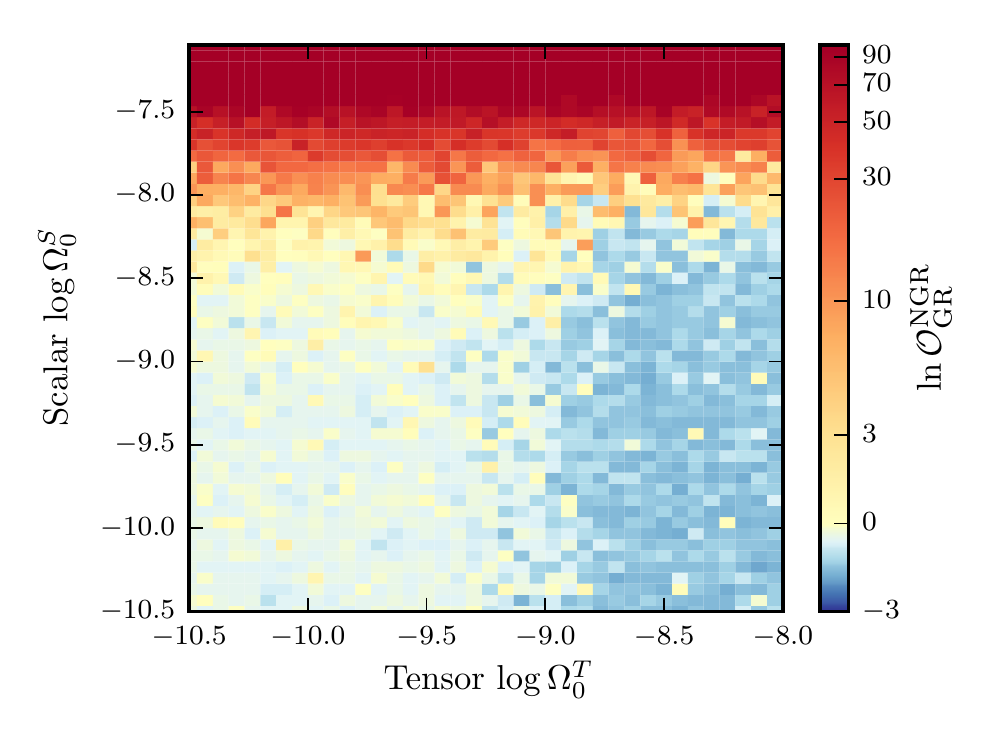}
  \caption{
As above, but for simulated three-year observations with the joint Advanced LIGO-Virgo network at design sensitivity.
Despite the inclusion of Advanced Virgo, the sensitivity of this three-detector network is nearly identical to that of Advanced LIGO alone.
  }
\label{STbayesHLV}
\end{figure*}

So far we have considered only cases of pure tensor, vector, or scalar polarization.
Plausible alternative theories of gravity, however, would typically predict a mixed background of multiple polarization modes.
How does our Bayesian machinery handle a background of mixed polarization?
To answer this question, we will investigate backgrounds of mixed tensor and scalar polarization.
Figure \ref{STbayes} shows values of $\OddsSN$ and $\OddsGR$ (left and right-hand sides, respectively) as a function of the amplitude of each polarization.
While we allow the amplitudes to vary, we fix the tensor and scalar slopes to $\alpha_T = 2/3$ (as predicted for binary black hole backgrounds) and $\alpha_S = 0$.

In the left side Fig. \ref{STbayes}, the recovered values of $\ln\OddsSN$ simply trace contours of total energy.
Thus the \textit{detectability} of a mixed background depends only on its total measured energy, rather than its polarization content.
Meanwhile, three distinct regions are observed in the right-hand subplot.
First, for small tensor and scalar amplitudes ($\log\Omega_0^T\lesssim\blue{-9.0}$ and $\log\Omega_0^S\lesssim\blue{-8.5}$), we obtain $\ln\OddsGR\approx\blue{-0.4}$.
In this region, the mixed background simply cannot be detected and so we recover the slight Occam's bias towards the GR hypothesis as noted above.
Secondly, for small scalar and large tensor amplitudes ($\log\Omega_0^T \gtrsim \blue{-9.0}$), the recovered odds ratios decrease to $\ln\OddsGR\approx\blue{-1.4}$.
This corresponds to the detection of the tensor component alone; the decrease in odds ratios is the same behavior previously seen in Figs. \ref{histograms} and \ref{oddsScatter}.
Finally, when $\Omega_0^S$ is large, the scalar component is detectable and the recovered $\ln\OddsGR$ increases rapidly to large, positive values.
The threshold value of $\Omega_0^S$ at which $\ln\OddsGR$ becomes positive shows only little dependence on the amplitude of any tensor background which might also be present.
When $\Omega_0^T$ is small, for instance, scalar amplitudes of size $\log\Omega_0^S\gtrsim\blue{-7.9}$ are required to preference the NGR model.
When $\Omega_0^T$ is large, this requirement increases only slightly to $\log\Omega_0^S\gtrsim\blue{-7.8}$.
Thus, we should expect Advanced LIGO to be able to both detect and \textit{identify as non-tensorial} a flat scalar background of amplitude $\log\Omega_0^S \gtrsim\blue{-8}$, regardless of the presence of an additional tensor component.

It should be pointed out that positive $\log\OddsGR$ indicates only that there exists evidence for alternative polarizations.
From the odds ratio alone we cannot infer which specific polarizations -- vector and/or scalar -- are present in the background.
While we found above that Advanced LIGO can identify mixed tensor-scalar backgrounds as non-tensorial when $\log\Omega^S_0\gtrsim-8$, this \textit{does not} imply that we can successfully identify the scalar component as such, only that our measurements are not consistent with tensor polarization alone (see Sect. \ref{peSection}).

\newtext{
The future addition of new gravitational wave detectors will extend the reach of stochastic searches and help to break degeneracies between backgrounds of different polarizations.
This expansion recently began with the completion of Advanced Virgo, which joined Advanced LIGO during its O2 observing run in August 2017 \cite{Acernese2015,GW170814}.
It is therefore interesting to investigate how the introduction of Advanced Virgo will improve the above results.
}
Given detectors indexed by $i\in\{1,2,...\}$, the total SNR of a stochastic background is the quadrature sum of SNRs from each detector pair \cite{Allen1999}:
	\begin{equation}
	\SNR^2 = \sum_{i} \sum_{j>i} \SNR^2_{ij},
	\end{equation}
where each $\SNR_{ij}$ is computed following Eq. \eqref{snr}.
Naively, the SNR with which a background is observed is expected to increase as $\SNR\propto\sqrt{N}$, where $N$ is the total number of available detector pairs (three in the case of the Advanced LIGO-Virgo network).
However, both the Hanford-Virgo and Livingston-Virgo pairs exhibit reduced sensitivity to the stochastic background due to their large physical separations.
This fact is reflected in their respective overlap reduction functions, which are a factor of several smaller in magnitude than the Hanford-Livingston overlap reduction functions (see Fig. \ref{orfPlot}).

Given three independent detector pairs (and hence three independent measurements at each frequency), one can in principle directly solve for the unknown tensor, vector, and scalar contributions to the background in each frequency bin \cite{Nishizawa2009,Nishizawa2010,Nishizawa2013,Romano2016}.
This component separation scheme can be performed without resorting to a model for the stochastic energy-density spectrum.
However, frequency-by-frequency component separation is unlikely to be successful using the LIGO-Virgo network, due to the large uncertainties in the measured background at each frequency.
Instead, when considering joint Advanced LIGO-Virgo observations we will again apply the Bayesian framework introduced above, leveraging measurements made at many frequencies in order to constrain the power-law amplitude and slope of each polarization mode.

To quantify the extent to which Advanced Virgo aids in the detection of the stochastic background, we again consider simulated observations of a mixed tensor (slope $\alpha_T=2/3$) and scalar (slope $\alpha_S=0$) background, this time with a three-detector Advanced LIGO-Virgo network.
Our Bayesian formalism is easily extended to accommodate the case of multiple detector pairs; details are given in Appendix \ref{modelConstruction}.
The odds ratios obtained from our simulated Advanced LIGO-Virgo observations are shown in Fig. \ref{STbayesHLV} for various tensor and scalar amplitudes.
The inclusion of Advanced Virgo yields no clear improvement over the Advanced LIGO results in Fig. \ref{STbayes}.
Due to its large distance from LIGO, Advanced Virgo does not contribute more than a small fraction of the total observed SNR.
As a result, the combined Hanford-Livingston-Virgo network both detects (as indicated with $\OddsSN$) and identifies (via $\OddsGR$) the scalar background component with virtually the same sensitivity as the Hanford-Livingston network alone.

\section{PARAMETER ESTIMATION ON MIXED BACKGROUNDS}
\label{peSection}

\begin{table*}[ht!]
\caption{Stochastic background parameters used for each case study presented.
For each case, the vector amplitude is set to zero.
Also shown are the odds ratios computed for each simulated observation.}
\label{peTable}
\setlength{\tabcolsep}{3pt}
\begin{tabular}{l | r r r r | r r | r r }
\hline
\hline
\multirow{2}{*}{Case} & \multirow{2}{*}{$\log\Omega^T_0$} & \multirow{2}{*}{$\alpha_T$} & \multirow{2}{*}{$\log\Omega^S_0$} & \multirow{2}{*}{$\alpha_S$} & \multicolumn{2}{c|}{H1-L1} & \multicolumn{2}{c}{H1-L1-V1} \\
&&&&& $\ln\mathcal{O}^\textsc{sig}_\textsc{n}$        &        $\ln\mathcal{O}^\textsc{ngr}_\textsc{gr}$     &        $\ln\mathcal{O}^\textsc{sig}_\textsc{n}$        &        $\ln\mathcal{O}^\textsc{ngr}_\textsc{gr}$ \\
\hline
1. Noise       &        -    &        -    &        -    &        -    &        -1.1   &        -0.4  &        -1.1   &        -0.4 \\
2. Tensor       &        -8.78  &        0.67   &        -    &       -    &        8.4    &        -1.4   &        8.8    &        -1.4 \\
3. Tensor+Scalar       &        -8.48  &        0.67   &        -7.83  &        0.0    &        193.5   &        16.1    &        197.3   &        19.3  \\
\hline
\hline
\end{tabular}
\end{table*}

Parameter estimation will be the final step in a search for a stochastic background of generic polarization.
If a gravitational-wave background is detected (as inferred from $\OddsSN$), how well can Advanced LIGO constrain the properties of the background?
Alternatively, if no detection is made, what upper limits can Advanced LIGO place on the background amplitudes of each polarization mode?
We investigate these questions through three case studies: an observation of pure Gaussian noise, a standard tensor stochastic background, and a background of mixed tensor and scalar polarizations.
The simulated background parameters used for each case are listed in Table \ref{peTable}.

When performing model selection above, the odds ratios $\OddsSN$ and $\OddsGR$ were constructed by independently allowing for each combination of tensor, vector, and scalar modes (see Appendix \ref{modelConstruction}).
Parameter estimation, meanwhile, must be performed in the context of a specific background model.
For the case studies below, we will adopt the broadest possible hypothesis, allowing for all three polarization modes (the TVS hypothesis in Appendix \ref{modelConstruction}).
This choice will allow us to place simultaneous constraints on the presence of tensor, vector, and scalar polarizations in the stochastic background.
Parameter estimation is achieved using \multinest, which returns samples drawn from the measured posterior distributions.

There are several key subtleties that must be understood when interpreting the parameter estimation results presented below.
First, whereas standard tensor upper limits are conventionally defined with respect to a single, fixed slope \cite{Aasi2014,TheLIGOScientificCollaboration2016b}, we will quote amplitude limits obtained \textit{after} marginalization over spectral index.
This approach concisely combines information from the entire posterior parameter space to offer a single limit on each polarization considered.
As a result, however, our simulated upper limits presented here should not be directly compared to those from standard searches for tensor backgrounds.
Secondly, parameter estimation results are contingent upon the choice of a specific model.
While we will demonstrate parameter estimation results under our TVS hypothesis (see Appendix \ref{modelConstruction}), other hypotheses may be better suited to answering other experimental questions.
For example, if we were specifically interested in constraining scalar-tensor theories (which \textit{a priori} do not allow vector polarizations), we would instead perform parameter estimation under the TS hypothesis.
And if our goal was to perform a standard stochastic search for a purely tensor-polarized background, we would restrict to the T hypothesis.
Although these various hypotheses all contain an analogous parameter $\Omega^T_0$, the resulting upper limits on $\Omega^T_0$ will generically be different in each case.
In short, different experimental questions will yield different answers.

\subsection*{Case 1: Gaussian Noise}

\begin{figure*}
\centering
\includegraphics[width=1\textwidth]{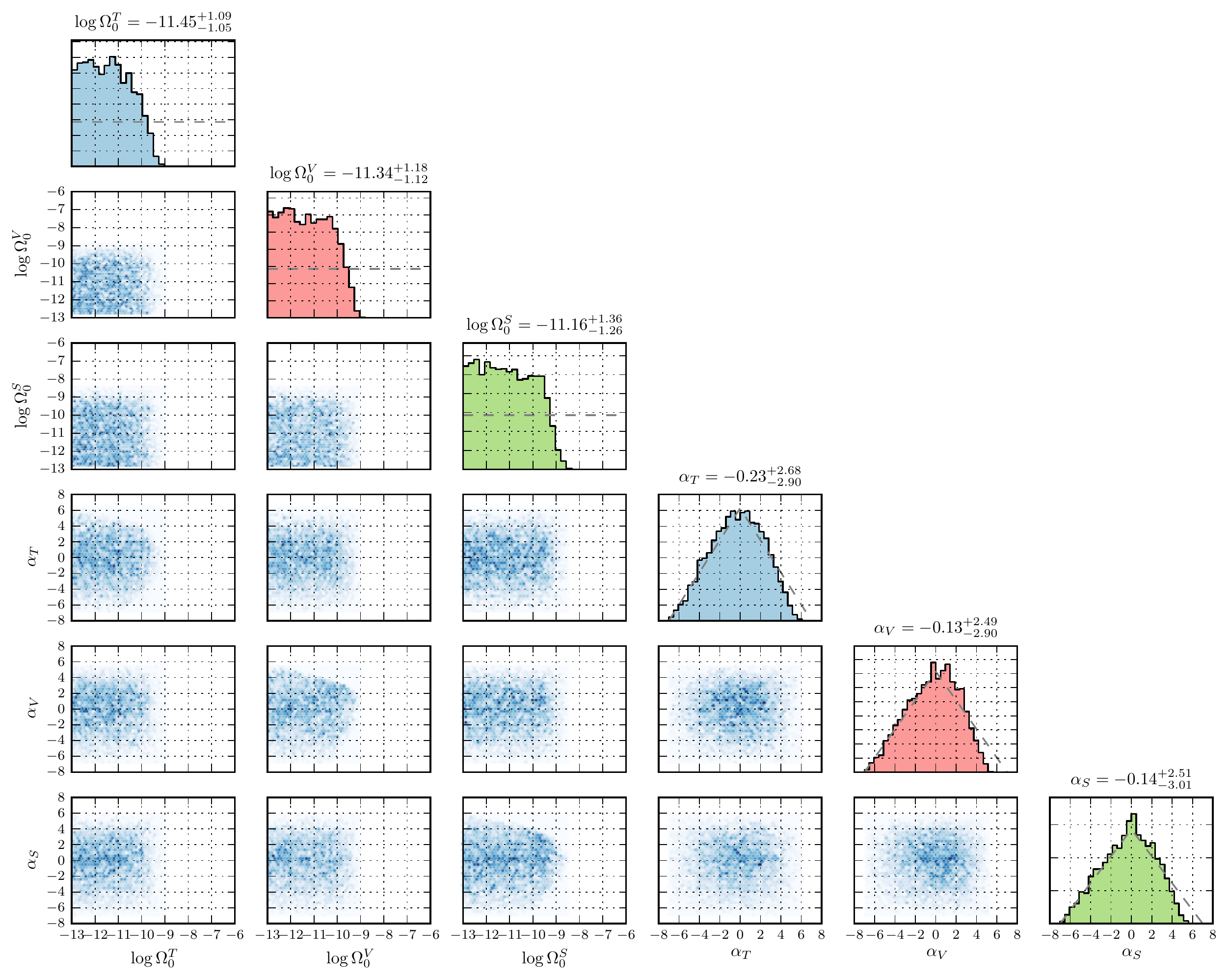}
\caption{
Posteriors obtained for a simulated Advanced LIGO observation of pure Gaussian noise (Case 1 in Table \ref{peTable}), under the TVS hypothesis.
The subplots along the diagonal show marginalized posteriors for the amplitudes and slopes of the tensor, vector, and scalar backgrounds (blue, red, and green, respectively), while the remaining subplots show the 2D posterior between each pair of parameters.
Each amplitude posterior is consistent with our lower prior bound, reflecting the non-detection of a stochastic background.
}
\label{NoisePE}
\end{figure*}
\begin{figure*}
\centering
\includegraphics[width=1\textwidth]{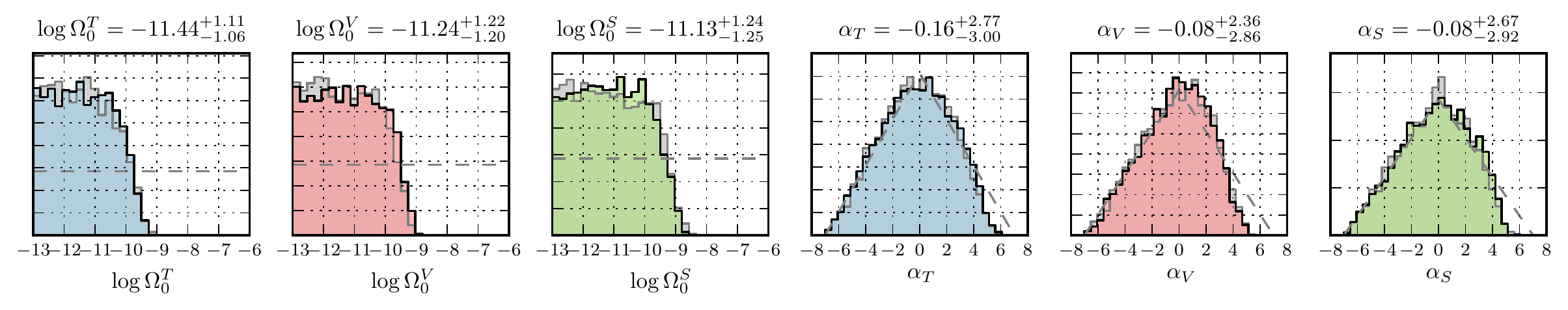}
\caption{
Marginalized amplitude and slope posteriors for the Gaussian noise observation in Fig. \ref{NoisePE}, after the additional inclusion of design-sensitivity Advanced Virgo.
For reference, the light grey histograms show the Advanced LIGO-only results from Fig. \ref{NoisePE}.
As above, dashed grey lines show the priors placed on each parameter.
We see that the inclusion of Advanced Virgo does not significantly affect the parameter estimation results.
}
\label{NoisePE_HLV}
\end{figure*}

First, we consider the case of pure noise, producing a simulated three-year observation of Gaussian noise at Advanced LIGO's design sensitivity.
The resulting TVS posteriors are shown in Fig. \ref{NoisePE}.
The colored histograms along the diagonal show the marginalized 1D posteriors for the amplitudes and slopes of the tensor, vector, and scalar components (blue, green, and red, respectively).
The priors placed on each parameter are indicated with a dashed grey curve.
Above each posterior we quote the median posterior value as well as $\pm34\%$ credible limits.
The remaining subplots illustrate the joint 2D posteriors between each pair of parameters.

For this simulated Advanced LIGO observation, we obtain $\log\OddsSN=\blue{-1.1}$, consistent with a null detection.
Accordingly, the posteriors on $\Omega^T_0$, $\Omega^V_0$, and $\Omega^S_0$ are each consistent with the lower bound of our amplitude prior (at $\log\Omega_\mathrm{Min} = -13$).
Meanwhile, the posteriors on spectral indices $\alpha_T$, $\alpha_V$, and $\alpha_S$ simply recover our chosen prior.
The 95\% credible upper limits on each amplitude are $\log\Omega^T_0 < \blue{-9.8}$, $\log\Omega^V_0 < \blue{-9.7}$, and $\log\Omega^S_0 < \blue{-9.3}$.

In Fig. \ref{NoisePE_HLV} we show the posteriors obtained if we additionally include design-sensitivity Advanced Virgo (incorporating simulated measurements for the HV and LV detector pairs).
For reference, the grey histograms show the posteriors from Fig. \ref{NoisePE} obtained by Advanced LIGO alone.
The Advanced LIGO-Virgo posteriors are virtually identical to those obtained from Advanced LIGO alone, with 95\% credible upper limits of $\log\Omega^T_0 < \blue{-9.9}$, $\log\Omega^V_0 < \blue{-9.6}$, and $\log\Omega^S_0 < \blue{-9.4}$.
In the case of a null-detection, then, the inclusion of Advanced Virgo does not notably improve the upper limits placed on the amplitudes of tensor, vector, and scalar backgrounds.

\subsection*{Case 2: Tensor Background}

\begin{figure*}
\centering
\includegraphics[width=1\textwidth]{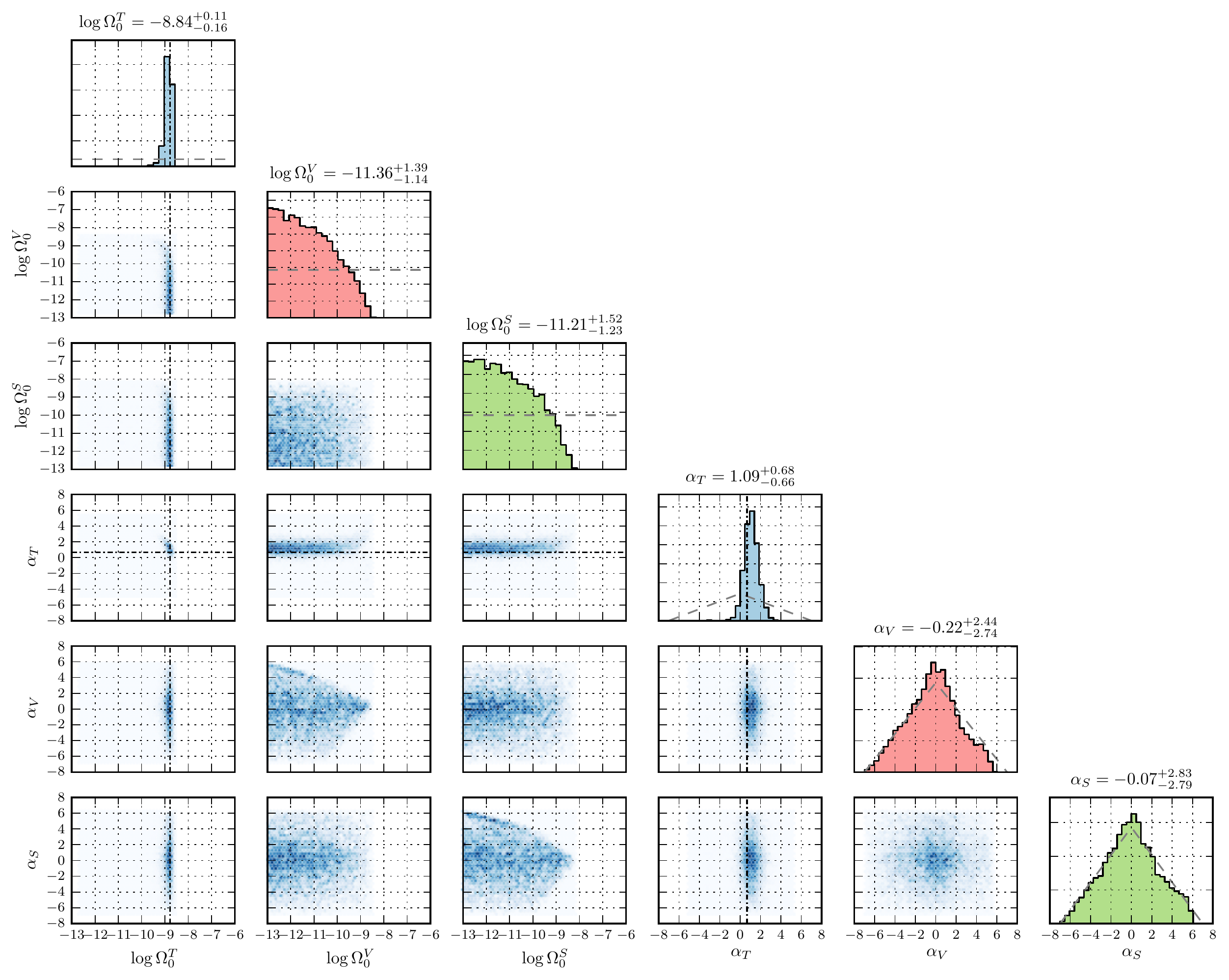}
\caption{
As in Fig. \ref{NoisePE}, for a simulated observation of a pure tensor background (Case 2 in Table \ref{peTable}).
The injected tensor amplitude and slope are indicated by dot-dashed black lines.
The tensor amplitude and slope posteriors are peaked about their true values.
The vector and scalar amplitude posteriors, meanwhile, are consistent with our lower prior bound.
}
\label{TensorPE}
\end{figure*}
\begin{figure*}
\centering
\includegraphics[width=1\textwidth]{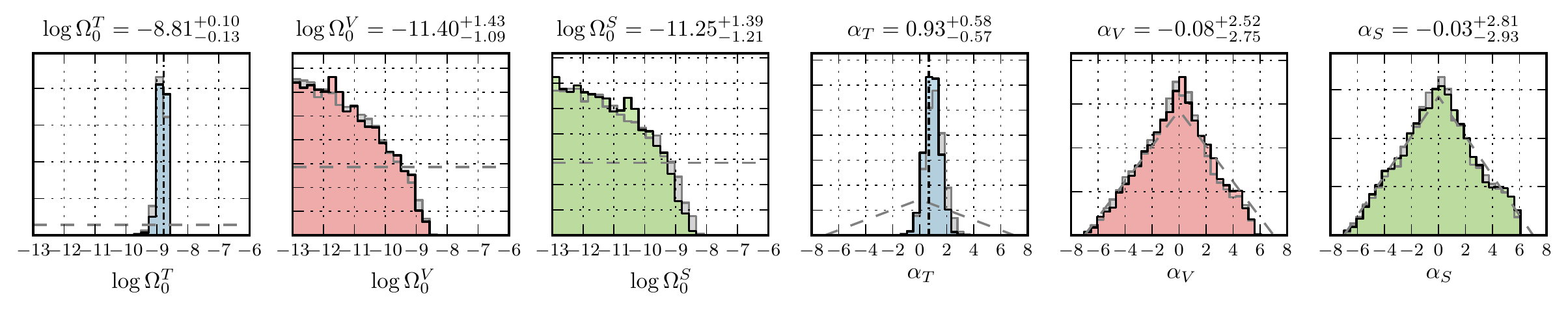}
\caption{
Marginalized amplitude and slope posteriors for the tensor background observation in Fig. \ref{TensorPE}, after the additional inclusion of design-sensitivity Advanced Virgo.
For reference, the light grey histograms show the Advanced LIGO-only results from Fig. \ref{TensorPE}.
The joint LIGO-Virgo parameter estimation yields a slightly tighter measurement of $\Omega^T_0$, as well as somewhat improved upper limits on $\Omega^V_0$ and $\Omega^S_0$.
}
\label{TensorPE_HLV}
\end{figure*}

\begin{figure*}
\centering
\includegraphics[width=1\textwidth]{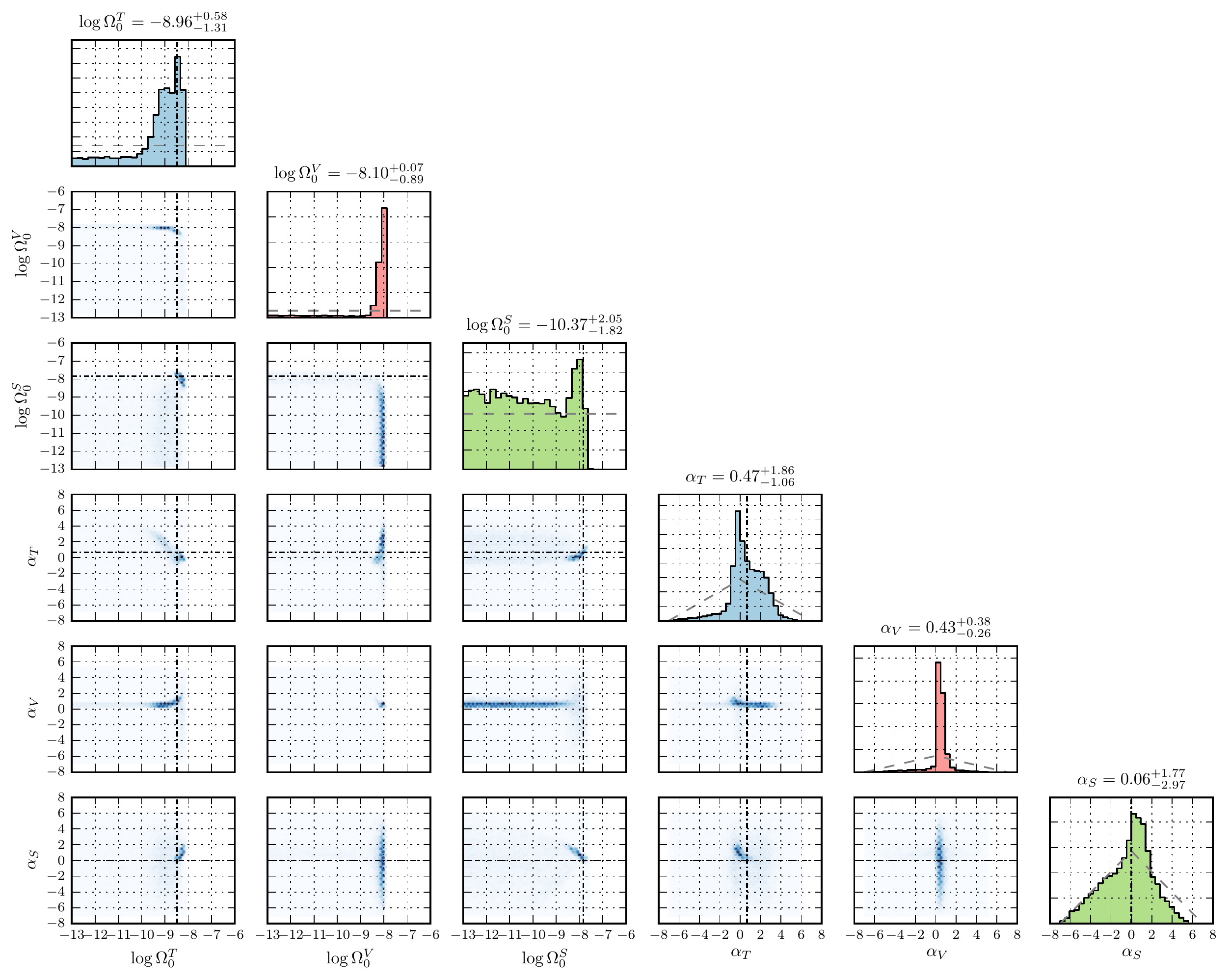} \\
\caption{
As in Figs. \ref{NoisePE} and \ref{TensorPE}, for a simulated observation of a mixed tensor and scalar background (Case 3 in Table \ref{peTable}).
While the $\Omega^T_0$ and $\Omega^S_0$ posteriors are locally peaked about the true values, much of the observed energy is mistaken for vector polarizations.
Thus Advanced LIGO alone is unable to break the degeneracy between tensor, vector, and scalar amplitudes.
}
\label{TensorFlatScalarPE}
\end{figure*}
\begin{figure*}
\centering
\includegraphics[width=1\textwidth]{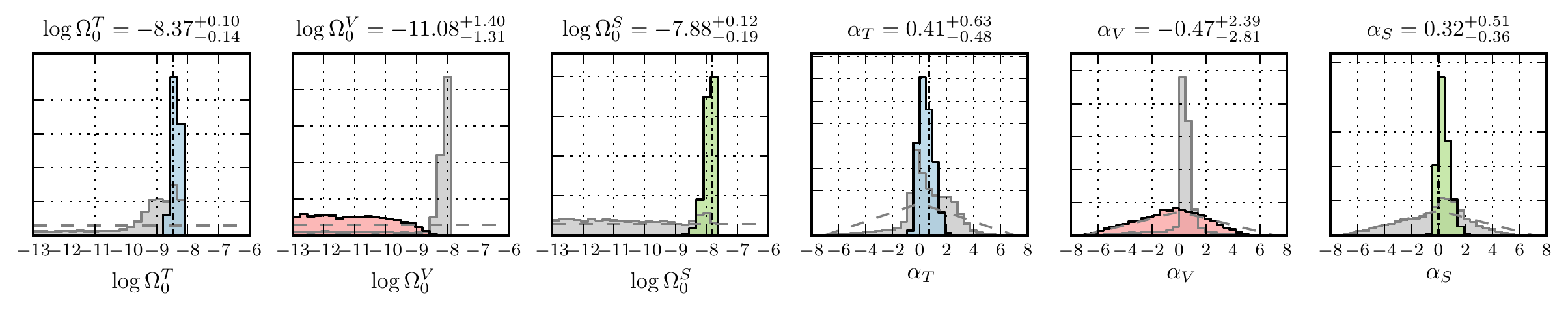} \\
\caption{
Marginalized amplitude and slope posteriors for the mixed tensor and scalar background observation in Fig. \ref{TensorFlatScalarPE}, after the additional inclusion of design-sensitivity Advanced Virgo.
For reference, the light grey histograms show the Advanced LIGO-only results from Fig. \ref{TensorFlatScalarPE}.
In contrast to the results in Fig. \ref{TensorFlatScalarPE}, the degeneracy between polarization modes is completely broken when including Advanced Virgo.
Thus, while Advanced Virgo does not particularly improve prospects for the detection of a mixed background, it can significantly improve our ability to perform parameter estimation on multiple modes simultaneously.
}
\label{TensorFlatScalarPE_HLV}
\end{figure*}

Next, we produce a simulated observation of a pure tensor background with amplitude $\log\Omega^T_0 = \blue{-8.78}$ and spectral index $\alpha_T = 2/3$.
The amplitude is chosen such that the background would be detected by Advanced LIGO with expected $\langle \SNRopt\rangle = 5$ after three years of observation at design-sensitivity.
The odds ratios obtained for this simulated observation are $\log\OddsSN=\blue{8.4}$ and $\log\OddsGR=\blue{-1.4}$, indicating a strong detection consistent with general relativity.

The corresponding parameter posteriors are shown in Fig. \ref{TensorPE}.
In this case, the injected parameter values are shown via dot-dashed black lines.
The $\log\Omega^T_0$ posterior is strongly peaked near the true value, with a central $68\%$ credible interval of $\blue{-9.0}\leq\log\Omega^T_0\leq\blue{-8.7}$ and a median value of $\log\Omega^T_0 = \blue{-8.8}$.
The vector and scalar amplitudes, in turn, are consistent with the lower bound on our prior, with 95\% credible upper limits of $\log\Omega^V_0 < \blue{-9.2}$ and $\log\Omega^S_0 < \blue{-9.0}$.

The parameter estimation results when additionally including Advanced Virgo are given in Fig. \ref{TensorPE_HLV}.
Once again, the grey histograms show parameter estimation results from Advanced LIGO alone.
Although Virgo does not improve our confidence in the detection, it \textit{can} serve to break degeneracies present between different polarization modes.
We begin to see this behavior in Fig. \ref{TensorPE_HLV}, in which the vector and scalar log-amplitude posteriors are pushed to smaller values in the joint LIGO-Virgo analysis.
When including Advanced Virgo, we obtain a marginally tighter $68\%$ credible interval of $\blue{-8.9}\leq\log\Omega^T_0\leq\blue{-8.7}$ on the tensor amplitude, and slightly improved upper limits of $\log\Omega^V_0 < \blue{-9.3}$ and $\log\Omega^S_0 < \blue{-9.2}$ on vector and scalar amplitudes.

\subsection*{Case 3: Tensor and Scalar Backgrounds}

As discussed above, most alternative theories of gravity would predict a stochastic background of mixed polarization.
For our final case study, we therefore consider a mixed background with both tensor ($\log\Omega^T_0 = \blue{-8.48}$ and $\alpha_T=2/3$) and scalar ($\log\Omega^S_0 = \blue{-7.83}$ and $\alpha_S = 0$) components.
The amplitudes are chosen such that each component is individually observable with $\langle \SNRopt\rangle = 10$ after three years of observation.
Analysis with \multinest yields odds ratios $\log\OddsSN = \blue{193.5}$ and $\log\OddsGR=\blue{16.1}$, representing an extremely loud detection with very strong evidence for the presence alternative polarizations.

The posteriors obtained for this data are shown in Fig. \ref{TensorFlatScalarPE}.
Despite the strength of the simulated stochastic signal, we see that parameter estimation results are dominated by degeneracies between the different polarization modes.
Although the tensor and scalar amplitude posteriors are locally peaked about their true values, much of the background's energy is misattributed to vector modes, illustrating that potential severe degeneracies persist even at high SNRs.
These degeneracies are exacerbated for backgrounds with small or negative spectral indices, as in the present case.
Such backgrounds preferentially weight low frequencies where the Advanced LIGO overlap reduction functions are all similar (see Fig. \ref{orfPlot}).
This example serves to illustrate that, while Advanced LIGO can likely identify the \textit{presence} of alternative polarizations through the odds ratio $\OddsGR$, Advanced LIGO alone is unable to determine which modes (vector or scalar) have been detected.

In contrast, the degeneracies in Fig. \ref{TensorFlatScalarPE} are completely broken with the inclusion of Advanced Virgo.
Whereas the $\Omega^V_0$ posterior is strongly peaked in Fig. \ref{TensorFlatScalarPE}, we see in Fig. \ref{TensorFlatScalarPE_HLV} that the posterior is instead entirely consistent with our lower prior bound when including Advanced Virgo.
The tensor and scalar amplitude posteriors, meanwhile, are each more strongly-peaked about their correct values and are now inconsistent with the lower amplitude bound.
Thus, while Advanced Virgo generally does not improve our ability to \textit{detect} a stochastic background, we see that it can significantly improve prospects for simultaneous parameter estimation of multiple polarizations.

\section{BROKEN TENSOR SPECTRA}
\label{brokenSection}

\begin{figure}
  \centering
  \includegraphics[width=0.48\textwidth]{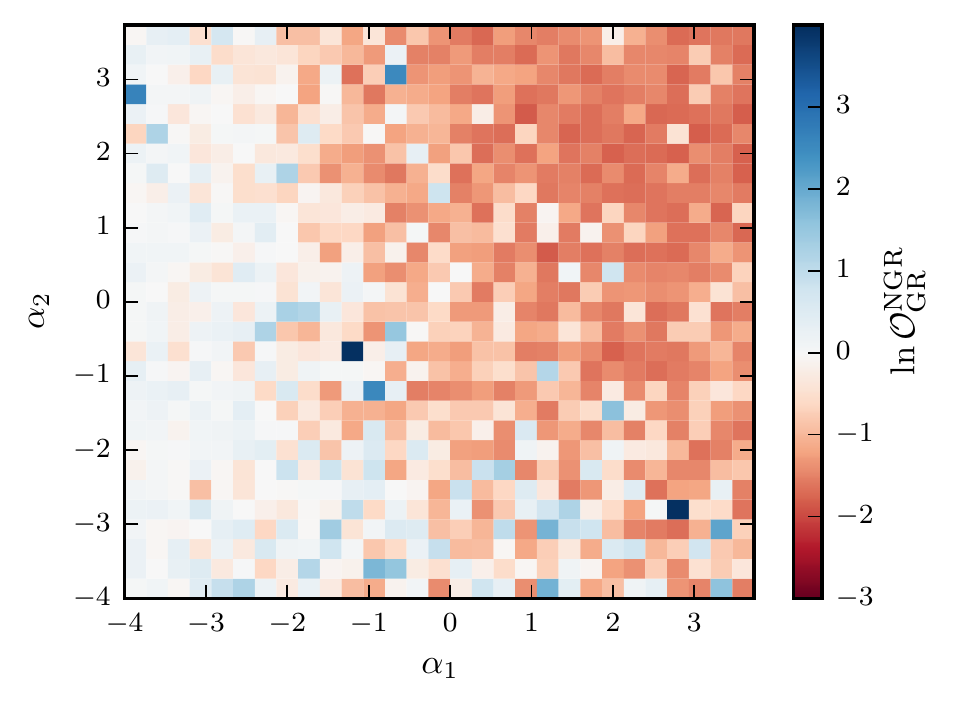}
  \caption{
Odds ratios $\OddsGR$ obtained for simulated Advanced LIGO observations of tensor-polarized broken power law backgrounds with energy density spectra given by Eq. \eqref{brokenEquation}.
The parameters $\alpha_1$ and $\alpha_2$ are the backgrounds' slopes below and above the ``knee" frequency $f_k$, which we take to be 30 Hz (in the center of the stochastic sensitivity band).
We scale the amplitude $\Omega_0$ of each background such that it is optimally detectable with $\langle \SNRopt\rangle = 5$ after the simulated observation period.
By design, these backgrounds are not well-described by single power laws, the form explicitly assumed in our search.
Despite this fact, we find that these backgrounds are \textit{not} systematically misclassified as containing vector or scalar polarization.
  }
  \label{brokenFig}
\end{figure}

The stochastic search presented here offers a means to search for alternative gravitational-wave polarizations in a nearly model-independent way.
Unlike direct searches for compact binary coalescences, our search makes minimal assumptions about the source and nature of the stochastic background.
We do, however, make one notable assumption: that the energy density spectra $\Omega^a(f)$ are well-described by power laws in the Advanced LIGO frequency band.
This is expected to be a reasonable approximation for most predicted astrophysical sources of gravitational waves.
The backgrounds expected from stellar-mass binary black holes \cite{Callister2016}, core-collapse supernovae \cite{Crocker2015}, and rotating neutron stars \cite{Regimbau2008,Wu2013,Talukder2014}, for instance, are all well-modelled by power laws in the Advanced LIGO band.
It may be, however, that the stochastic background is in fact \textit{not} well-described by a single power law.
This may be the case if, for instance, the background is dominated by high-mass binary black holes, an excess of systems at high redshift, or previously-unexpected sources of gravitational waves \cite{Callister2016}.

Given that our search allows only for power-law background models, how would we interpret a non-power-law background?
In particular, if the stochastic background is purely tensorial (obeying general relativity) but is not well-described by a power-law, would our search mistakenly claim evidence for alternative polarizations?

To investigate this question, we consider simulated Advanced LIGO observations of pure tensor backgrounds described by broken power laws:
	\begin{equation}
	\Omega^T(f) = \begin{cases}
		\Omega_0 \left(\frac{f}{f_k}\right)^{\alpha_1} & (f<f_k) \\
		\Omega_0 \left(\frac{f}{f_k}\right)^{\alpha_2} & (f\geq f_k).
		\end{cases}
	\label{brokenEquation}
	\end{equation}
Here, $\Omega_0$ is the background's amplitude at the ``knee frequency" $f_k$, while $\alpha_1$ and $\alpha_2$ are the slopes below and above the knee frequency, respectively.
We will set the knee frequency to $f_k=30\Hz$, placing the backgrounds' knees in the most sensitive band of the stochastic search.
The odds ratios $\OddsGR$ we obtain for these broken power laws are shown in Fig. \ref{brokenFig} as a function of the two slopes $\alpha_1$ and $\alpha_2$.
Each simulation assumes three years of observation at design-sensitivity, and the amplitudes $\Omega_0$ are scaled such that each background has expected $\langle \SNRopt \rangle = 5$ after this time.
Any trends in Fig. \ref{brokenFig} are therefore due to the backgrounds' spectral shapes rather than their amplitudes.

If tensor broken power laws are indeed misclassified by our search, we should expect large, positive $\ln\OddsGR$ values in Fig. \ref{brokenFig}.
Instead, we see that broken power laws are \textit{not} systematically misclassified.
When $\alpha_1$ and $\alpha_2$ are each positive, we recover $\ln\OddsGR\approx-1.5$, correctly classifying backgrounds as tensorial despite the fact that they are not described by power laws.
When $\alpha_1<0$, meanwhile, we recover odds ratios scattered about $\ln\OddsGR\approx 0$.
This simply reflects the fact that when $\alpha_1$ is negative the majority of a background's SNR is collected at low frequencies where Advanced LIGO's tensor, vector, and scalar overlap reduction functions are degenerate.
In such a case we do not show preference for either model over the other.
Note that we find $\ln\OddsGR\approx0$ even along the line $\alpha_2=\alpha_1$ (for $\alpha_1<0$), where the background \textit{is} described by a single power law.

We expect broken power laws to be most problematic when $\alpha_1>0$ and $\alpha_2<0$; in this case a background's SNR is dominated by a small frequency band around the knee itself.
This would be the case if, for instance, the stochastic background were dominated by unexpectedly massive binary black hole mergers \cite{Callister2016}.
Figure \ref{brokenFig} does suggest a larger scatter in $\log\OddsGR$ for such backgrounds.
Even in this region, however, there is not a systematic bias towards larger values of $\OddsGR$, and the largest recovered odds ratios have $\log\OddsGR\lesssim2.5$, well below the level required to confidently claim evidence for the presence of alternative polarizations.

Despite the fact that we assume purely power-law models for the stochastic energy-density spectra, our search appears reasonably robust against broken power law spectra that are otherwise purely tensor-polarized.
In particular, in order to be mistakenly classified by our search, a tensor stochastic background would have to emulate the pattern of positive and negative cross-power associated with the vector and/or scalar overlap reduction functions (see, for instance, Fig. \ref{exampleBackgrounds}).
This is simply not easy to do without a pathological background.
While we have demonstrated this only for Advanced LIGO, we find similarly robust results for three-detector Advanced LIGO-Virgo observations.

Nevertheless, when interpreting odds ratios $\OddsGR$ it should be kept in mind that the true stochastic background may deviate from a power law.
Even if a broken tensor background is not misclassified in our analysis, the \textit{parameter estimation} results we obtain would likely be incorrect (another example of so-called ``stealth bias'').
It should be pointed out, though, that our analysis is not fundamentally restricted to power-law models.
While we adopt power-law models here for computational simplicity, our analysis can be straightforwardly expanded in the future to include more complex models for the stochastic energy-density spectrum.

\section{DISCUSSION}

\newtext{
The direct detection of gravitational waves by Advanced LIGO and Virgo has opened up new and unique prospects for testing general relativity.
One such avenue is the search for vector and scalar gravitational-wave polarizations, predicted by some alternative theories of gravity but prohibited by general relativity.
Observation of vector or scalar polarizations in the stochastic background would therefore represent a clear violation of general relativity.
While the first preliminary measurements have recently been made of the polarization of GW170814, our ability to study the polarization of transient gravitational-wave signals is currently limited by the number and orientation of current-generation detectors.
In contrast, searches for long-duration sources like the stochastic background offer a promising means of directly measuring gravitational-wave polarizations with existing detectors.
}

In this paper, we explored a procedure by which Advanced LIGO can detect or constrain the presence of vector and scalar polarizations in the stochastic background.
In Sect. \ref{nonGRBackgrounds}, we found that a stochastic background dominated by alternative polarization modes may be missed by current searches optimized only for tensor polarizations.
In particular, backgrounds of vector and scalar polarizations with large, positive slopes may take up to ten times as long to detect with current methods, relative to a search optimized for alternative polarizations.
In Sect. \ref{bayesianSearch}, we therefore proposed a Bayesian method with which to detect a generically-polarized stochastic background.
This method relies on the construction of two odds ratios (see Appendix \ref{modelConstruction}).
The first serves to determine if a stochastic background has been detected, while the second quantifies evidence for the presence of alternative polarizations in the background.
This search has the advantage of being entirely generic; it is capable of detecting and identifying stochastic backgrounds containing any combination of gravitational-wave polarizations.
With this method, we demonstrated flat scalar-polarized backgrounds of amplitude $\Omega^S_0\approx\blue{2\times10^{-8}}$ can be confidently identified as non-tensorial with Advanced LIGO.

In Sect. \ref{peSection}, we then considered the ability of Advanced LIGO to perform simultaneous parameter estimation on tensor, vector, and scalar components of the stochastic background.
After three years of observation at design sensitivity, Advanced LIGO will be able to limit the amplitudes of tensor, vector, and scalar polarizations to $\Omega^T_0 < \blue{1.6\times10^{-10}}$, $\Omega^V_0<\blue{2.0\times10^{-10}}$, and $\Omega^S_0<\blue{5.0\times10^{-10}}$, respectively, at 95\% credibility.
If, however, a stochastic background of mixed polarization is detected, Advanced LIGO alone cannot precisely determine the parameters of the tensor, vector, and/or scalar components simultaneously due to large degeneracies between modes.

We also considered how the addition of Advanced Virgo to the Hanford-Livingston network affects the search for alternative polarizations.
In Sect. \ref{bayesianSearch}, we found that addition of Advanced Virgo does not particularly increase our ability to detect or identify backgrounds of alternative polarizations.
However, we found in Sect.\ref{peSection} that Advanced Virgo \textit{does} significantly improve our ability to perform parameter estimation on power-law backgrounds, breaking the degeneracies that plagued the Hanford-Livingston analysis.

Relative to other modeled searches for gravitational waves, the stochastic search described here has the advantage of being nearly model-independent.
We have, however, made one large assumption: that the tensor, vector, and scalar energy-density spectra are well-described by power laws in the Advanced LIGO band.
Finally, in Sect. \ref{brokenSection} we explored the implications of this assumption, asking the question: would tensor backgrounds \textit{not} described by power laws be mistaken for alternative polarizations in our search?
We found that our proposed Bayesian method is reasonably robust against this possibility.
In particular, even pure tensor backgrounds with sharply-broken power law spectra are not systematically misidentified by our search.

The non-detection of alternative polarizations in the stochastic background may yield interesting experimental constraints on extended theories of gravity.
Meanwhile, any experimental evidence \textit{for} alternative polarizations in the stochastic background would be a remarkable step forward for experimental tests of gravity.
Of course, if future stochastic searches do yield evidence for alternative polarizations, careful study would be required to verify that this result is not due to unmodeled effects like non-Gaussianity or anisotropy in the stochastic background \cite{2017PhRvL.118l1102A,Meacher2014,Martellini2014,Thrane2013,Cornish2015}.
Comparison to polarization measurements of other long-lived sources like rotating neutron stars \cite{2015PhRvD..91h2002I,Max} will additionally aid in the interpretation of stochastic search results.

Several future developments may further improve the ability of ground-based detectors to detect alternative polarization modes in the stochastic background.
First, the continued expansion of the ground-based detector network will improve our ability to both resolve the stochastic background and accurately determine its polarization content.
Secondly, while we presently assume that the stochastic background is Gaussian, the background contribution from binary black holes is expected to be highly non-Gaussian \cite{TheLIGOScientificCollaboration2016}.
Future stochastic searches may therefore be aided by the development of novel data analysis techniques optimized for non-Gaussian backgrounds \cite{Thrane2013,Martellini2014,Cornish2015}.

\acknowledgements{
We would like to thank Thomas Dent, Gregg Harry, Joe Romano, and Alan Weinstein for their careful reading of this manuscript, as well as many members of the LIGO-Virgo Collaboration Stochastic Backgrounds working group for helpful comments and conversation.
T. C. and M. I. are members of the LIGO Laboratory, supported by funding from the U. S. National Science Foundation.
LIGO was constructed by the California Institute of Technology and Massachusetts Institute of Technology with funding from the National Science Foundation and operates under cooperative agreement PHY-0757058.
N.C. is supported by NSF grant PHY-1505373.
The work of A.M. was supported in part by the NSF grant PHY-1204944 at the University of Minnesota.
M.S. is partially supported by STFC (UK) under the research grant ST/L000326/1.
E.T. is supported through ARC FT150100281 and CE170100004.
This paper carries the LIGO Document Number LIGO-P1700059 and King's College London report number KCL-PH-TH/2017-25.
}

\begin{appendix}

\section{OVERLAP REDUCTION FUNCTIONS}
\label{orfAppendix}

The sensitivity of a two-detector network to a stochastic gravitational-wave background is quantified by the overlap reduction function \cite{Christensen1992,Allen1999}
	\begin{equation}
	\gamma(f) \propto \sum_A \int e^{2\pi i f \hatbf{\Omega} \cdot \Delta \mathbf{x}/c} F^A_1(\hatbf\Omega)
		F^A_2(\hatbf\Omega) d\mathbf{\Omega},
	\label{orfEq}
	\end{equation}
where $\Delta \mathbf{x}$ is the displacement vector between detectors, $c$ is the speed of light, and $F^A_{1/2}(\hatbf{\Omega})$ are the antenna patterns describing the response of each detector to gravitational-waves of polarization $A$ propagating from the direction $\hatbf{\Omega}$.
The overlap reduction function is effectively the sky-averaged product of the two detectors' antenna patterns, weighted by the additional phase accumulated as a gravitational wave propagates from one site to the other.

In the standard stochastic search, the summation in Eq. \eqref{orfEq} is taken over the tensor plus and cross polarizations.
When extending the stochastic search to generic gravitational-wave polarizations, we now must consider three separate overlap reduction functions for the tensor, vector, and scalar modes \cite{Nishizawa2009}:
	\begin{equation}
	\begin{aligned}
	\gamma_T(f) &= \frac{5}{8\pi} \sum_{A=\{+,\times\}} \int e^{2\pi i f \hatbf{\Omega} \cdot \Delta \mathbf{x}/c} F^A_1(\hatbf\Omega)
		F^A_2(\hatbf\Omega) d\mathbf{\Omega}, \\
	\gamma_V(f) &= \frac{5}{8\pi} \sum_{A=\{x,y\}} \int e^{2\pi i f \hatbf{\Omega} \cdot \Delta \mathbf{x}/c} F^A_1(\hatbf\Omega)
		F^A_2(\hatbf\Omega) d\mathbf{\Omega}, \\
	\gamma_S(f) &= \frac{5}{12\pi} \sum_{A=\{b,l\}} \int e^{2\pi i f \hatbf{\Omega} \cdot \Delta \mathbf{x}/c} F^A_1(\hatbf\Omega)
		F^A_2(\hatbf\Omega) d\mathbf{\Omega}.
	\end{aligned}
	\label{orfIntegrals}
	\end{equation}
We normalize these functions such that $\gamma_T(f) = 1$ for coincident, co-aligned detectors; detectors that are rotated or separated relative to one another have $\gamma_T(f) < 1$.
The amplitudes of $\gamma_V(f)$ and $\gamma_S(f)$, meanwhile, express relative sensitivities to vector and scalar backgrounds.

Note that the normalization of $\gamma_S(f)$ differs from that of Nishizawa \textit{et al.} in Ref. \cite{Nishizawa2009}.
This difference is due to Nishizawa \textit{et al.}'s definition of the longitudinal polarization tensor as
	\begin{equation}
	\underaccent{\sim}{\hatbf{e}}^l = \sqrt{2} \hatbf{\Omega} \otimes \hatbf{\Omega},
	\end{equation}
rather than the more common
	\begin{equation}
	\hatbf{e}^l = \hatbf{\Omega} \otimes \hatbf{\Omega}
	\end{equation}
(to distinguish between these two conventions, the quantities adopted by Nishizawa \textit{et al.} will be underscored with tildes).
As a consequence, Nishizawa \textit{et al.} obtain a longitudinal antenna pattern
	\begin{equation}
	\underaccent{\sim}{F}^l(\hatbf{\Omega}) = \frac{1}{\sqrt{2}} \sin^2\theta \cos2\phi,
	\end{equation}
which differs by a factor of $\sqrt{2}$ from the conventional form
	\begin{equation}
	F^l(\hatbf{\Omega}) = \frac{1}{2}\sin^2\theta \cos2\phi.
	\end{equation}
Correspondingly, the quantity $\underaccent{\sim}{\Omega}^l(f)$ defined by Nishizawa \textit{et al.} is actually half of the canonical energy density in longitudinal gravitational waves:
	\begin{equation}
	\underaccent{\sim}{\Omega}^l(f)= \frac{1}{2} \Omega^l(f).
	\end{equation}

While each overlap reduction function may be calculated numerically via Eq. \eqref{orfIntegrals}, they may also be analytically expanded in terms of spherical Bessel functions \cite{Allen1999,Nishizawa2009}.
See Ref. \cite{Nishizawa2009} for definitions of the tensor, vector, and scalar overlap reduction functions in this analytic form.
Note, however, that these definitions follow Nishizawa \textit{et al.}'s normalization convention as discussed above; the analytic expression given for $\gamma_S(f)$ must be divided by 3 to match our Eq. \eqref{orfIntegrals}.

\section{OPTIMAL SIGNAL-TO-NOISE RATIO}
\label{snrAppendix}

Searches for the stochastic background rely on measurements $\hat C(f)$ of the cross-power between two detectors.
As discussed in Sect. \ref{nonGRBackgrounds}, the expectation value and variance of $\hat C(f)$ are given by Eqs. \eqref{Y} and \eqref{sigma2}, respectively.
Here, we derive the optimal broadband signal-to-noise ratio [Eq. \eqref{optimalSNR}], which combines a spectrum of cross-correlation measurements into a single detection statistic.

Given a measured spectrum $\hat C(f)$ and associated uncertainties $\sigma^2(f)$, a single broadband statistic may be formed via the weighted sum
	\begin{equation}
	\hat C = \frac{\sum_f \hat C(f) w(f)/\sigma^2(f)}{\sum_f w(f)/\sigma^2(f)},
	\end{equation}
where $w(f)$ is a set of yet-undefined weights.
The mean and variance of $\hat C$ are
	\begin{equation}
	\langle \hat C \rangle = \frac{\sum_f \gamma_a(f) \Omega^a(f) w(f)/\sigma^2(f)}{\sum_f w(f)/\sigma^2(f)},
	\end{equation}
and
	\begin{equation}
	\sigma^2 = \frac{\sum_f w^2(f)/\sigma^2(f)}{\left(\sum_f w(f)/\sigma^2(f)\right)^2},
	\end{equation}
where $\gamma_a(f) \Omega^a(f)$ denotes summation $\sum_a \gamma_a(f) \Omega^a(f)$ over polarization modes ${a\in\{T,V,S\}}$.

We define a broadband signal-to-noise ratio by $\text{SNR} = \hat C/\sigma$.
In the limit $df\to0$, this quantity may be written
	\begin{equation}
	\label{suboptSNR}
	\text{SNR} = \frac{\left(\hat C\,|\, w\right)}{ \sqrt{\left(w\,|\,w\right)}},
	\end{equation}
where we have substituted Eq. \eqref{sigma2} for $\sigma^2(f)$ and made use of the inner product defined in Eq. \eqref{innerProduct}.
The expected SNR is maximized when the chosen weights are equal to the true background, such that $w(f) = \gamma_a(f)\Omega^a_\textsc{gw}(f)$.
In this case, the optimal expected SNR of the stochastic background becomes
	\begin{equation}
	\langle\text{SNR}_\text{opt}\rangle = \sqrt{\left( \gamma_a\Omega^a_\textsc{gw} \,|\, \gamma_b\Omega^b_\textsc{gw} \right)}.
	\end{equation}

\section{ODDS RATIO CONSTRUCTION}
\label{modelConstruction}

Here, we describe the construction of the odds ratios $\OddsSN$ and $\OddsGR$ introduced in Sect. \ref{bayesianSearch}.
Given data $\hat C(f)$, the Bayesian evidence for some hypothesis $\mathcal{A}$ with parameters $\theta_A$ is defined
	\begin{equation}
	\label{evidence}
	P(\hat C | \mathcal{A}) = \int \mathcal{L}(\hat C | \theta_A,\mathcal{A}) \pi(\theta_A|\mathcal{A}) d\theta_A.
	\end{equation}
Here, the likelihood $\mathcal{L}(\hat C |\theta_A,\mathcal{A})$ gives the conditional probability of the measured data under hypothesis $\mathcal{A}$ for fixed parameter values,
while $\pi(\theta_A | \mathcal{A})$ is the prior probability set on these parameters.
When selecting between two such hypotheses $\mathcal{A}$ and $\mathcal{B}$, we may define an odds ratio
	\begin{equation}
	\label{odds}
	\mathcal{O}^\mathcal{A}_\mathcal{B} = \frac{
			P(\hat C | \mathcal{A})
			}{
			P(\hat C | \mathcal{B})
			}
			\frac{\pi(\mathcal{A})}{\pi(\mathcal{B})}.
	\end{equation}
The first factor in Eq. \eqref{odds}, called the Bayes factor, is the ratio between the Bayesian evidences for hypotheses $\mathcal{A}$ and $\mathcal{B}$.
The second term, meanwhile, is the ratio between the prior probabilities $\pi(\mathcal{A})$ and $\pi(\mathcal{B})$ assigned to each hypothesis.

To construct odds ratios for our stochastic background analysis, we will first need the likelihood $\mathcal{L}(\{\hat C\}|\theta,\mathcal{A})$ of a measured cross-power spectrum under model $\mathcal{A}$ with some parameters $\theta$.
In the presence of Gaussian noise, the likelihood of measuring a specific $\hat C(f)$ within a single frequency bin is \cite{Allen1999,Mandic2012,Callister2016}
	\begin{equation}
	\label{likelihoodA}
	\mathcal{L}\left(\hat C(f) | \theta,\mathcal{A}\right) \propto
		\exp \left( - \frac{ \left[\hat C(f) - \gamma_a(f)\Omega^a_\mathcal{A} (\theta; f) \right]^2 }{ 2\sigma^2(f)} \right),
	\end{equation}
with variance $\sigma^2(f)$ given by Eq. \eqref{sigma2}.
Here, $\Omega^a_\mathcal{A}(\theta;f)$ is our model for the energy-density spectrum under hypothesis $\mathcal{A}$ and with parameters $\theta$, evaluated at the given frequency $f$.
The full likelihood $\mathcal{L}(\{\hat C\}|\theta,\mathcal{A})$ for a spectrum of cross-correlation measurements is the product of the individual likelihoods in each frequency bin:
	\begin{equation}
	\label{likelihood}
	\begin{aligned}
	\mathcal{L} (\{\hat C\} | \theta,\mathcal{A}) 
		&\propto \prod_f \mathcal{L}(\hat C(f) | \theta,\mathcal{A}) \\
		&= \mathcal{N} \exp \left[ -\frac{1}{2}
			\left( \hat C-\gamma_a\Omega^a_\mathcal{A} \,|\, \hat C-\gamma_b\Omega^b_\mathcal{A}\right) \right],
	\end{aligned}
	\end{equation}
where $\mathcal{N}$ is a normalization coefficient and we have used the inner product defined by Eq. \eqref{innerProduct}.

As discussed in Sect. \ref{bayesianSearch}, we will seek to detect and characterize a generic stochastic background via the construction of two odds ratios: $\OddsSN$, which indicates whether a background of any polarization is present, and $\OddsGR$, which quantifies evidence for the presence of alternative polarization modes.
First consider $\OddsSN$.
Under the noise hypothesis (N), we assume that no signal is present [such that $\Omega_\textsc{n}^a(f) = 0$].
From Eq. \eqref{likelihood}, the corresponding likelihood is simply
	\begin{equation}
	\label{noiseLikelihood}
	\mathcal{L}(\{\hat C\} | \mathrm{N} ) = \mathcal{N} \exp \left[ - \frac{1}{2} \left( \hat C \,|\, \hat C\right)\right].
	\end{equation}

The signal hypothesis (SIG) is somewhat more complex.
The signal hypothesis is ultimately the union of seven distinct sub-hypotheses that together describe all possible combinations of tensor, vector, and scalar polarizations \cite{2012PhRvD..85h2003L,Max}.
To understand this, first define a ``TVS'' hypothesis that allows for the simultaneous presence of tensor, vector, and scalar polarization.
In this case, we will model the stochastic energy-density spectrum as a sum of three power laws
	\begin{equation}
	\Omega_\textsc{tvs}(f) = \Omega^T_0 \left(\frac{f}{f_0}\right)^{\alpha_T}
		+ \Omega^V_0 \left(\frac{f}{f_0}\right)^{\alpha_V}
		+ \Omega^S_0 \left(\frac{f}{f_0}\right)^{\alpha_S},
	\end{equation}
with free parameters $\Omega^a_0$ and $\alpha_a$ setting the amplitude and spectral index of each polarization sector.
The priors on these parameters are given by Eqs. \eqref{ampPrior} and \eqref{alphaPrior} below.

In defining the TVS hypothesis, we have made the explicit assumption that tensor, vector, and scalar radiation are each present.
This is not the only possibility, of course.
A second distinct hypothesis, for instance, is that only tensor and vector polarizations exist.
This is our ``TV" hypothesis.
We model the corresponding energy spectrum as
	\begin{equation}
	\Omega_\textsc{tv}(f) = \Omega^T_0 \left(\frac{f}{f_0}\right)^{\alpha_T}
		+ \Omega^V_0 \left(\frac{f}{f_0}\right)^{\alpha_V}.
	\end{equation}
In a similar fashion, we must ultimately define seven such hypotheses, denoted TVS, TV, TS, VS, T, V, and S, to encompass all combinations of tensor, vector, and scalar gravitational-wave backgrounds.
Our complete signal hypothesis is given by the union of these seven sub-hypotheses \cite{2012PhRvD..85h2003L,Max}.
For each signal sub-hypothesis, we adopt the log-amplitude and slope priors given below in Eqs. \eqref{ampPrior} and \eqref{alphaPrior}.

\begin{figure}
  \centering
  \includegraphics[width=0.48\textwidth]{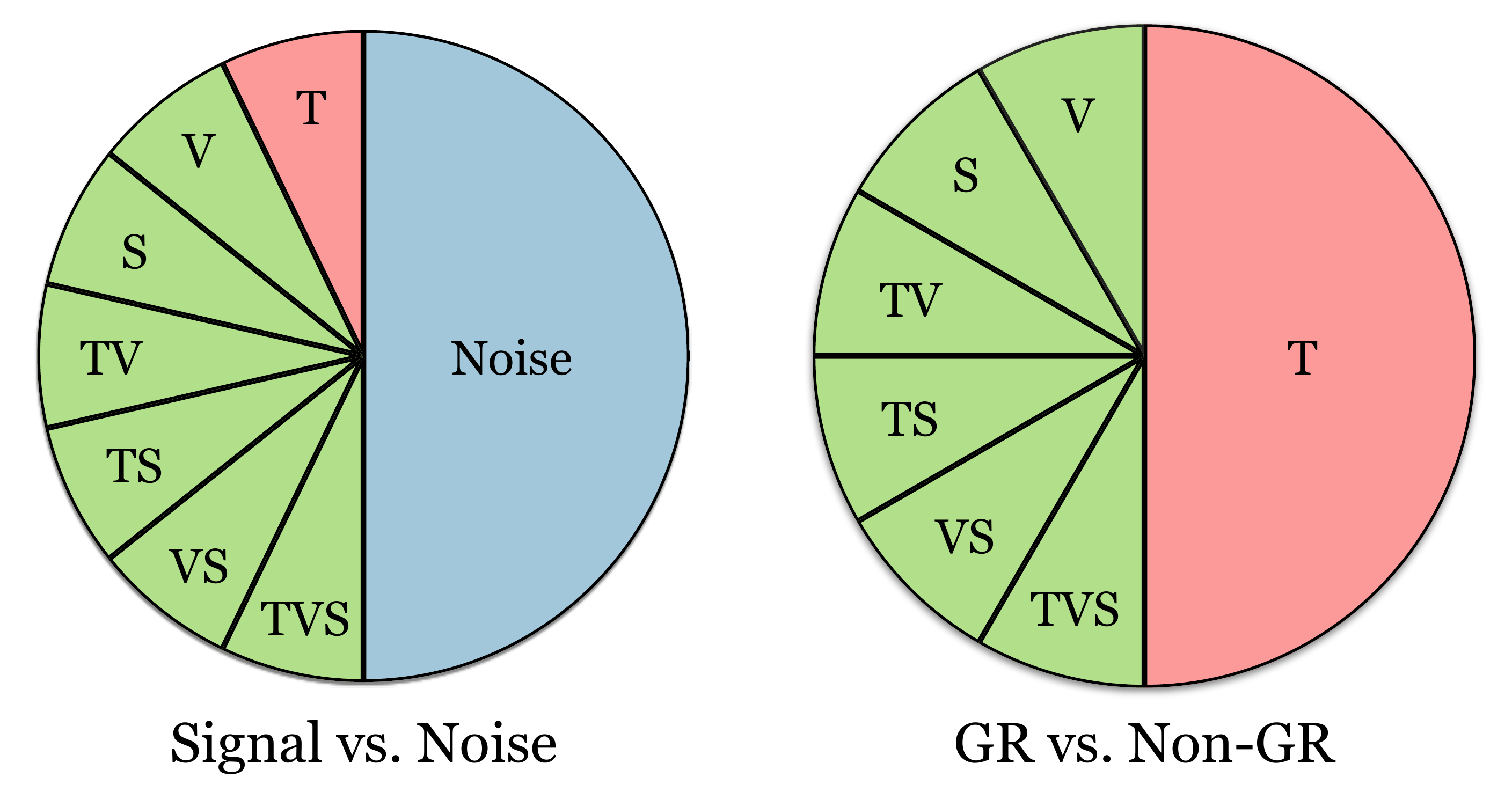}
  \caption{
Illustration of the prior odds assigned to models and sub-hypotheses in the hierarchical Bayesian search for non-GR stochastic backgrounds.
When constructing $\OddsSN$, we assign equal prior probability to the noise and signal models, as well as equal probability to the seven signal sub-hypotheses $\{\mT,...,\mTVS\}$.
Similarly, when constructing $\OddsGR$, we give equal probability to the non-GR and GR models and identically weight the six non-GR sub-hypotheses $\{\mV,...,\mTVS\}$.
  }
  \label{priorFig}
\end{figure}

Each of the signal sub-hypotheses are logically independent \cite{Max,2012PhRvD..85h2003L}, and so the odds ratio $\OddsSN$ between signal and noise hypotheses is given by the sum of odds ratios between the noise hypothesis and each of the seven signal sub-hypotheses:
	\begin{equation}
	\OddsSN = \hspace{2mm} \sum_{\mathclap{\mathcal{A}\in\{\mT,\mV,\mS,...\}}}
		\hspace{2mm}\mathcal{O}^\mathcal{A}_\textsc{n}.
	\end{equation}
As illustrated in Fig. \ref{priorFig}, we assign equal prior probability to the signal and noise hypotheses.
Within the signal hypothesis, we weight each of the signal sub-hypotheses equally, such that the prior odds between e.g. the T and N hypothesis is $\pi(\text{T})/\pi(\text{N})=1/7$.
We note that our choice of prior probabilities is not unique; there may exist other valid choices as well.
Our analysis can easily accommodate different choices of prior weight.

The odds ratio $\OddsGR$ is constructed similarly.
In this case, we are selecting between the hypothesis that the stochastic background is purely tensor-polarized (GR), or the hypothesis that additional polarization modes are present (NGR).
The GR hypothesis is identical to our tensor-only hypothesis T from above:
	\begin{equation}
	\Omega_\textsc{gr}(f) = \Omega^T_0 \left(\frac{f}{f_0}\right)^{\alpha_T}.
	\end{equation}
The NGR hypothesis, on the other hand, will be the union of the six signal sub-hypotheses that are inconsistent with general relativity: V, S, TV, TS, VS, and TVS.
The complete odds ratio between NGR and GR hypothesis is then
	\begin{equation}
	\OddsGR = \hspace{2mm} \sum_{\mathclap{\mathcal{A}\in\{\mV,\mS,\mTV,...\}}} \mathcal{O}^\mathcal{A}_\mT.
	\end{equation}
As shown in Fig. \ref{priorFig}, we have assigned equal priors to the GR and NGR hypotheses as well as identical priors to the six NGR sub-hypotheses.

In computing the odds ratios $\OddsSN$ and $\OddsGR$, we also need priors for the various parameters governing each model for the stochastic background.
In the various energy-density models presented above, we have defined two classes of parameters: amplitudes $\Omega^a_0$ and spectral indices $\alpha_a$ of the background's various polarization components.
For each amplitude parameter, we will use the prior 
	\begin{equation}
	\label{ampPrior}
	\pi(\Omega_0) \propto \begin{cases}
		1/\Omega_0 & \left( \Omega_\text{Min} \leq \Omega_0 \leq \Omega_\text{Max} \right) \\
		0 & \left(\text{Otherwise}\right)
		\end{cases}.
	\end{equation}
This corresponds to a uniform prior in the log-amplitudes between $\log\Omega_\text{Min}$ and $\log\Omega_\text{Max}$.
In order for this prior to be normalizable, we cannot let it extend all the way to $\Omega_\text{Min} = 0$ ($\log\Omega_\text{Min} \to -\infty$).
Instead, we must choose a finite lower bound.
While this lower bound is somewhat arbitrary, our results depend only weakly on the specific choice of bound \cite{Max}.
In this paper, we take $\Omega_\text{Min} = 10^{-13}$, an amplitude that is indistinguishable from noise with Advanced LIGO.
Our upper bound, meanwhile, is $\Omega_\text{Max} = 10^{-6}$, consistent with upper limits placed by Initial LIGO and Virgo~\cite{Aasi2014}.

We adopt a \textit{triangular} prior on $\alpha$, centered at zero:
	\begin{equation}
	\pi(\alpha) = \begin{cases}
			\frac{1}{\alpha_\text{Max}} \left(1-\frac{|\alpha|}{\alpha_\text{Max}}\right)
				& \left( |\alpha| \leq\alpha_\text{Max}\right)\\
			0 & \left(\text{Otherwise}\right)
		\end{cases}.
	\label{alphaPrior}
	\end{equation}
This prior has several desirable properties.
First, it captures a natural tendency for spectral index posteriors to peak symmetrically about $\alpha = 0$.
As a result, our $\alpha$ posteriors reliably recover this prior in the absence of informative data (see Fig. \ref{NoisePE}, for example).
Second, this prior preferentially weights shallower energy-density spectra.
This quantifies our expectation that the stochastic background's energy density be distributed somewhat uniformly across logarithmic frequency intervals (at least in the LIGO band), rather than entirely at very high or very low frequencies.

Alternatively, Eq. \eqref{alphaPrior} can be viewed as corresponding to equal priors on the background strength at two different frequencies.
To understand this, first note that $\alpha$ may be written as a function of background amplitudes $\Omega_0$ and $\Omega_1$ at two frequencies $f_0$ and $f_1$:
	\begin{equation}
	\alpha(\Omega_1,\Omega_2) = \frac{\log\left(\Omega_1/\Omega_0\right)}{\log\left(f_1/f_0\right)}.
	\end{equation}
The prior probability of a particular slope $\alpha$ is equal to the probability of drawing any two amplitudes $\Omega_1$ and $\Omega_2$ satisfying $\log(\Omega_1/\Omega_2) = \alpha \log(f_1/f_2)$.
This is given by the convolution
	\begin{equation}
	\label{priorConvolution}
	\pi(\alpha) = \int \pi(\log\Omega_1) \pi(\log\Omega_0 = \log\Omega_1 - \alpha \log(f_1/f_0)) d\log\Omega_1.
	\end{equation}
For simplicity, we will set $f_1 = 10 f_0$ (such that $\log(f_1/f_0) = 1$) and place identical log-uniform priors [Eq. \eqref{ampPrior}] on each amplitude.
Under these assumptions, Eq. \eqref{priorConvolution} yields Eq. \eqref{alphaPrior}.

In Sects. \ref{bayesianSearch} and \ref{peSection}, we additionally considered the performance of the three-detector Advanced LIGO-Virgo network.
The Bayesian framework considered here is easily extended to accommodate multiple detector pairs.
The three LIGO and Virgo detectors allow for the measurement of three cross-correlation spectra: $\hat C^\textsc{hl}(f)$, $\hat C^\textsc{hv}(f)$, and $\hat C^\textsc{lv}(f)$.
In the small signal limit ($\Omega^a(f) \ll 1$), the correlations between these measurements vanish at leading order, and so the three baselines can be treated as statistically independent \cite{Allen1999}.
We can therefore factorize the joint likelihood for the three sets:
	\begin{equation}
	\begin{aligned}
	\mathcal{L}(\{&\hat C^\textsc{hl}, \hat C^\textsc{hv}, \hat C^\textsc{lv} \} | \theta,\mathcal{A}) \\
		&= \mathcal{L} (\{\hat C^\textsc{hl}\}|\theta,\mathcal{A})\,
			\mathcal{L} (\{\hat C^\textsc{hv}\}|\theta,\mathcal{A})\,
			\mathcal{L} (\{\hat C^\textsc{lv}\}|\theta,\mathcal{A}) \\
		&= \mathcal{N} \exp \Bigl\{ -\frac{1}{2} \Bigl[
			\begin{aligned}[t]
				& \left( \hat C^\textsc{hl}-\gamma^\textsc{hl}_a\Omega^a_\mathcal{A} \,|\, 
					\hat C^\textsc{hl}-\gamma^\textsc{hl}_b\Omega^b_\mathcal{A}\right) \\
				&+ \left( \hat C^\textsc{hv}-\gamma^\textsc{hv}_a\Omega^a_\mathcal{A} \,|\, 
					\hat C^\textsc{hv}-\gamma^\textsc{hv}_b\Omega^b_\mathcal{A}\right) \\
				&+ \left( \hat C^\textsc{lv}-\gamma^\textsc{lv}_a\Omega^a_\mathcal{A} \,|\,
					\hat C^\textsc{lv}-\gamma^\textsc{lv}_b\Omega^b_\mathcal{A}\right) \Bigr]\Bigr\},
			\end{aligned}
	\end{aligned}
	\end{equation}
substituting likelihoods of the form \eqref{likelihood} for each pair of detectors.
Note that we have explicitly distinguished between the overlap reduction functions for each baseline, and $\mathcal{N}$ is again a normalization constant.
Other than the above change to the likelihood, all other details of the odds ratio construction is unchanged when including three detectors.

\section{EVALUATING BAYESIAN EVIDENCES WITH \texttt{MULTINEST}}
\label{multinestAppendix}

\begin{figure}
  \centering
  \includegraphics[width=0.48\textwidth]{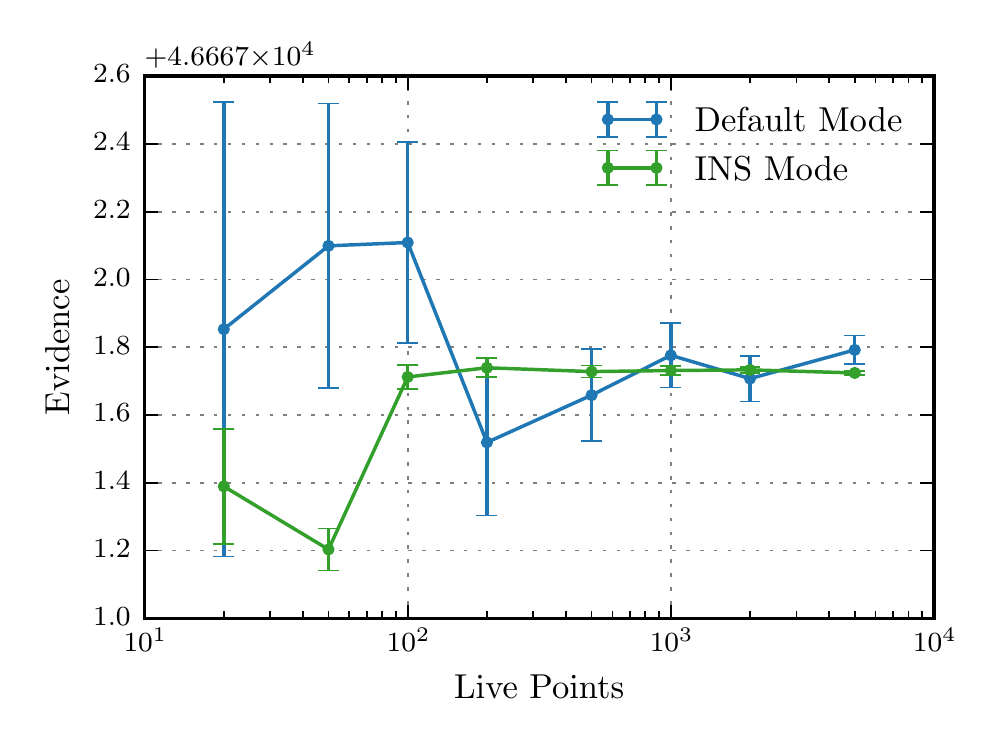}
  \caption{
\texttt{MultiNest} Bayesian evidences for a single simulated SGWB observation as a function of the number of live points chosen.
For the simulated data, we assume a tensor-polarized background (with $\Omega_0^T = 2\times10^{-8}$ and $\alpha_T = 2/3$) observed for one year with design-sensitivity Advanced LIGO, and compute evidence using the T model (see Sect. \ref{bayesianSearch}).
Results are shown for both \texttt{MultiNest}'s Default and INS modes; also shown are the error estimates provided by each mode.
To compute the results presented in this paper, we used $n=2000$ live points.
  }
  \label{livePoints}
\end{figure}

\begin{figure}
  \centering
  \includegraphics[width=0.48\textwidth]{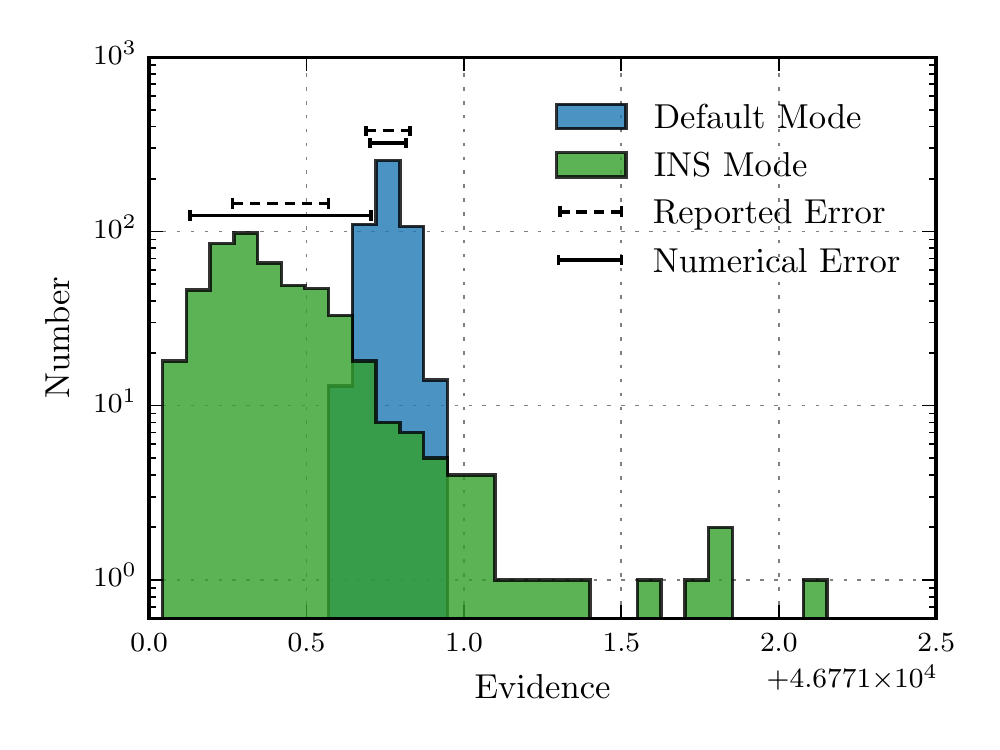}
  \caption{
Histograms of \texttt{MultiNest} evidences (for the TVS model; see Sect. \ref{bayesianSearch}) obtained by evaluating a single simulated data set 500 times in both the Default and INS modes.
To generate the simulated data, we assume a one-year observation of a tensor background ($\Omega_0^T = 2\times10^{-8}$ and $\alpha_T = 2/3$) with design-sensitivity Advanced LIGO.
The dashed error bars show the mean $68\%$ confidence interval reported by each method, while the solid error bars show the true $68\%$ confidence interval computed from the evidence distributions.
  }
  \label{multinestErrors}
\end{figure}

Here we summarize details associated with using \texttt{MultiNest} to evaluate Bayesian evidences for various models of the stochastic background.
The \texttt{MultiNest} algorithm allows for several user-defined parameters, including the number $n$ of live points used to sample the prior volume and the sampling efficiency $\epsilon$, which governs acceptance rate of new proposed live points (see e.g. Ref. \cite{Feroz2009} for details).
\texttt{MultiNest} also provides the option to run in Default or Importance Nested Sampling (INS) modes, each of which use different methods to evaluate evidences \cite{Feroz2013}.

To set the number of live points, we investigated the convergence of \texttt{MultiNest}'s evidence estimates with increasing values of $n$.
For a single simulated observation of a tensorial background (with amplitude $\Omega_0^T = 2\times10^{-8}$ and slope $\alpha_T=2/3$), for instance, Fig. \ref{livePoints} shows the recovered evidence for the T hypothesis (see Appendix \ref{modelConstruction} above) as a function of $n$, using both the Default (blue) and INS modes (green).
The results are reasonably stable for $n\gtrsim1000$; we choose $n=2000$ live points.
Meanwhile, our recovered evidence estimates do not exhibit noticeable dependence on the sampling efficiency; we choose the recommended values $\epsilon=0.3$ for evidence evaluation and $\epsilon=0.8$ for parameter estimation \cite{Feroz2009}.

In addition to computing Bayesian evidences, \texttt{MultiNest} also returns an estimate of the numerical error associated with each evidence calculation.
See, for instance, the error bars in Fig. \ref{livePoints}.
To gauge the accuracy of these error estimates, we construct a single simulated Advanced LIGO observation of a purely-tensorial stochastic background (again with $\Omega_0^T = 2\times10^{-8}$ and $\alpha_T=2/3$).
We then use \texttt{MultiNest} to compute the corresponding TVS evidence 500 times, in both Default and INS modes.
The resulting distributions of evidences are shown in Fig. \ref{multinestErrors}.
The dashed error bars show the averaged $\pm1\sigma$ intervals reported by \texttt{MultiNest}, while the solid bars show the $\pm1\sigma$ intervals obtained manually from the distributions.
We see that the errors reported by \texttt{MultiNest}'s Default mode appear to accurately reflect the numerical error in the evidence calculation, while the errors reported by the INS mode are underestimated by a factor of $\sim2$.

Additionally, Fig. \ref{multinestErrors} illustrates several systematic differences between the Default and INS results.
First, Default mode appears significantly more precise than INS mode, giving rise to a much narrower distribution of evidences.
Not only is the INS evidence distribution wider, but it exhibits a large tail extending several units in evidence above the mean.
We find that similarly long tails also appear for other pairs of injected signals and recovered models.
For this reason, we choose to use \texttt{MultiNest}'s Default mode in all evidence calculations.
Typical numerical errors in Default mode are of order $\delta(\mathrm{evidence}) \sim 0.1$, and so the uncertainty associated with a log-odds ratio is $\delta(\ln\mathcal{O})\sim \sqrt{2} \delta(\mathrm{evidence})$, again of order $0.1$.
Additionally, we see that the peaks of the Default and INS distributions do not coincide.
In general, the peaks of evidence distributions from the Default and INS modes lie $\sim0.3$ units apart.
Thus there may be additional systematic uncertainties in a given evidence calculation.
However, as long as we consistently use one mode or the other (in our case, Default mode), any uniform systematic offset in the evidences will simply cancel when we ultimately compute a log-odds ratio.

\end{appendix}

\bibliography{main.bbl}

\end{document}